\documentclass[%
preprint,
showpacs,preprintnumbers,
showkeys,
 amsmath,amssymb,
prd
]{revtex4-1}

\usepackage{graphicx}
\usepackage{dcolumn}
\usepackage{bm}
\usepackage{verbatim}
\usepackage{setspace}
\usepackage{float}

\usepackage[colorlinks=true
,anchorcolor=blue
,filecolor=blue
,linkcolor=blue
,menucolor=blue
,pagecolor=blue
,linktocpage=true
,pdfproducer=medialab
,pdfa=true
,citecolor=darkred
,urlcolor=darkred
,pdfborder={0 0 0}
]{hyperref}

\usepackage{subcaption}
\usepackage[usenames,dvipsnames]{xcolor} 
\usepackage[compat=1.1.0]{tikz-feynhand}
\usepackage{tikz-feynman}
\tikzfeynmanset{compat=1.1.0}
\usepackage{feynmp}
\usepackage{tikzsymbols}
\usepackage{array}
\usepackage{pifont} 
\definecolor{titlecolor}{rgb}{0.6,0,0}
\definecolor{mightnightblue}{RGB}{25,25,112}
\definecolor{brown}{rgb}{0.59, 0.29, 0.0}
\definecolor{darkred}{rgb}{0.6,0,0}
\definecolor{linkcolor}{rgb}{0,0,0.5}
\usepackage[colorinlistoftodos]{todonotes}
\usepackage[normalem]{ulem}

\usepackage{bbold}

\definecolor{titlecolor}{rgb}{0.6,0,0}
\definecolor{mightnightblue}{RGB}{25,25,112}
\definecolor{brown}{rgb}{0.59, 0.29, 0.0}
\definecolor{darkred}{rgb}{0.6,0,0}
\definecolor{linkcolor}{rgb}{0,0,0.5}

\usepackage{natbib}

\begin{document}


\title{\boldmath \color{BrickRed} New parameterization and analysis for E6 inspired 331 model}

\author{Rena \c{C}ift\c{c}i\footnote{rena.ciftci@ege.edu.tr}}
 \affiliation{Department of Physics, Faculty of Science, Ege University, 35040 Bornova, Izmir, T\"{u}rkiye}
\author{Abbas Kenan \c{C}ift\c{c}i\footnote{kenan.ciftci@ieu.edu.tr}}%
\affiliation{Department of Physics, Faculty of Arts and Sciences, Izmir University of Economics, 35330 Bal\c{c}ova, Izmir, T\"{u}rkiye}%

\author{Oleg Popov\footnote{opopo001@ucr.edu, Corresponding author.}}
\affiliation{Department of Biology, Shenzhen MSU-BIT University,\\ 1, International University Park Road, Shenzhen 518172, China}%
\affiliation{Department of Physics, Korea Advanced Institute of Science and Technology, \\291 Daehak-ro, Yuseong-gu, Daejeon 34141, Republic of Korea}%
\date{\today\\ \vspace{2cm}}
\begin{abstract}
We present a new parameterization for $SU(3)_C\times SU(3)_L\times U(1)_X$ extension of the Standard Model, which is inspired by $E_6$ symmetry. New setup predicts all Cabibbo-Kobayashi-Maskawa mixing angles and quark masses, total of nine observable variables, within 1-3 standard deviations of the experimental values with a minimum number of input parameters. A detailed numerical analysis and correlations between input parameters and predicted quantities are presented. The best global fit benchmark point corresponds to $\chi^2\approx 0.7$ with $\forall\sigma<0.6$. Advantages of the new parameterization and future prospects are discussed as well.
\end{abstract}

\maketitle
%
\twocolumngrid
\begin{singlespace}
\tableofcontents
\end{singlespace}
\onecolumngrid


%
\section{Introduction}
\label{sec:intro}

In spite of its effectiveness in accurately explaining electromagnetic, strong and weak interactions, the Standard Model (SM) has significant unresolved issues, such as the large mass spectrum and hierarchy of fermions, the small quark mixing angles, the existence of three fermion families, CP violation, the origin of neutrino masses, dark matter etc. Many extensions of SM, have been investigated to address some of these issues. The so-called 331 models are one of the simplest extensions, which modify electroweak gauge group of SM from $SU(3)_C\otimes SU(2)_L \otimes U(1)_Y$ to a $SU(3)_C\otimes SU(3)_L\otimes U(1)_X$ symmetry (331, hereafter). At the beginning, these models were presented as a natural explanation for the observed number of fermion families in nature. Various variants of the 331 model have been studied in detail to date. This model can be made anomaly-free in various ways. The 331 model can be made anomaly-free within the family as in the SM, or in other variants, it can be made anomaly-free by using all 3 families. The second approach is very attractive as it may be a natural explanation for the number  of families being 3 in the SM. 

331 based model has been focus of many studies and motivated by solving problems in various phenomenological applications. For instance, works on 331 model have been applied in the flavour physics~\cite{Duy:2022qhy,Addazi:2022frt,Buras:2013dea,Buras:2012dp}, neutrino mass generation~\cite{Boucenna:2014ela,Tully:2000kk}, and other phenomenological issues~\cite{Singer:1980sw,Pisano:1992bxx,Frampton:1992wt,Reig:2016tuk,Long:1995ctv,CarcamoHernandez:2005ka,Liu:1993gy,Profumo:2013sca,CarcamoHernandez:2013krw,Fonseca:2016tbn}. Furthermore, in view of the well known W boson mass anomaly, which was reported recently by the Collider Detector at Fermilab (CDF) Collaboration taken at Tevatron particle accelerator~\cite{CDF:2022hxs}, possible connection between W mass anomaly and the super-symmetric variation of the 331 model was examined~\cite{Rodriguez:2022wix}. For recent works on 331 models please refer to \cite{CarcamoHernandez:2013zrj,CarcamoHernandez:2014wdl,CarcamoHernandez:2017cwi,Barreto:2017xix}. On the other hand, 331 based models can be viewed as a precursor to grand unification models at high energy scales~\cite{Deppisch:2016jzl,Kownacki:2017uyq,Kownacki:2018lkj}. Finally, 3311 model, which is an extended variation of the 331 model, has been investigated in conjunction with dark matter candidates and neutrino mass generation mechanism~\cite{Alves:2016fqe,Dong:2017zxo,Kang:2019sab,Leite:2019grf}.

The mass spectrum and the mixing of quarks are one of important unsolved problems of the particle physics. Experimental values of quark masses at the scale of mass of Z boson are listed as $m_d=2.67\pm 0.19$~MeV, $m_s= 53.16\pm 4.61$~MeV, $m_b= 2.839\pm 0.026$~GeV, $m_u=  1.23\pm0.21$~MeV, $m_c= 620\pm 17$~MeV, $m_t= 168.26\pm 0.75$~GeV~\cite{Fritzsch:2021ipb}, whereas masses in ref.~\cite{Zyla:2020zbs} are given at different scales. Experimental limits~\cite{Fritzsch:2021ipb} at $M_Z=91.1876$~GeV scale of Cabibbo-Kobayashi-Maskawa (CKM) mixing matrix is
\begin{align}
\label{eq:ckm}
{\small V^{W}_{\text{CKM}}} &
{\small
=\mkern-9mu\left(
\begin{matrix}
\left| V_{ud}\right|  &\left| V_{us}\right|   &\left| V_{ub}\right| \\
\left| V_{cd}\right|  &\left| V_{cs}\right|   &\left| V_{cb}\right| \\
\left| V_{td}\right|  &\left| V_{ts}\right|   &\left| V_{tb}\right|
\end{matrix}
\right)
}
{\small=\mkern-9mu \left( 
\begin{matrix}
0.97401\pm 0.00011 & 0.22650\pm 0.00048 & 0.00361^{+0.00011}_{-0.00009} \\
0.22636\pm 0.00048 & 0.97320\pm 0.00011 & 0.04053^{+0.00083}_{-0.00061} \\ 
0.00854^{+0.00023}_{-0.00016} & 0.03978^{+0.00082}_{-0.00060} & 0.999172^{+0.000024}_{-0.000035}
\end{matrix}
\right)
}
\end{align}

There are number of attempts to explain masses and relation of CKM matrix elements to them. Somehow parameterization of CKM quark and Pontecorvo-Maki-Nakagawa-Sakata (PMNS) neutrino mixing matrices have always been very intriguing problems of the particle physics. For example, Wolfenstein parameterization and its various extensions ~\cite{Dattoli:2013fp,Branco:1988ba} have gotten some attention. Triminimal and tribimaximal approaches have been also  very popular parameterization methods ~\cite{Rodejohann:2008ir,Dattoli:2013fp,He:2008td,Li:2008aa,Branco:1988ba}. Number of particle physicists used exponential ~\cite{Dattoli:2013fp,Branco:1988ba}, recursive ~\cite{Chaturvedi:2010qf}, re-phasing invariants ~\cite{Branco:1987mj} parameterizations and unified pameterization of quarks and lepton matrices ~\cite{Li:2005ir,Li:2008aa} also parameterization involving eigenvalues ~\cite{Chaturvedi:2002zi} can be listed for the mixing matrix parameterization. However, this incomplete list of parameterizations does not try to solve mixing in conjunction with mass hierarchy. Recent work on CKM and PMNS parametrization is given in Ref.~\cite{Ganguly:2022cbo}.

 The Democratic Mass Matrix (DMM) approach has been proposed mainly by H. Fritzsch ~\cite{HARARI1978459,FRITZSCH1979189,FRITZSCH1987391,FRITZSCH1990451,FRITZSCH1994290} for the SM. In this approach, all quarks with the same quantum number behave equally under weak interaction in the up and down sectors and they are indistinguishable before the symmetry breaking:

\begin{equation}
\mathcal{M}^{0}_{u}=h_u\left( \begin{array}{ccc}
1 & 1  &  1 \\
1 & 1  &  1 \\
1 & 1  &  1 
\end{array}
 \right) 
\end{equation}
and
\begin{equation}
\mathcal{M}^{0}_{d}=h_d\left( \begin{array}{ccc}
1 & 1  &  1 \\
1 & 1  &  1 \\
1 & 1  &  1 
\end{array}
\right).
\end{equation}
Since there is only one Higgs field in the SM, $h_u=h_d$ is expected. In this case, it is naturally expected that $m_b\cong m_t$. The reason why the masses of these quarks are not close to each other, and therefore the wrong answer of the above mass matrix approach, which is so natural, may be due to the fact that the SM is not the last and most basic theory. To circumvent this, it was proposed to extend the SM to 4 families ~\cite{Datta,CELIKEL1995257,Atag,CIFTCI.055001,CIFTCI.053006}. As can be easily seen in this model, the mass difference between the t and b quarks can be explained inherently. However, ATLAS and CMS data excluded the 4th family ~\cite{DJOUADI2012310,collaboration2013searches}. Although the 4th family vector quark is not excluded by the experimental data, this deviates from the  V-A theory, the natural approach of the SM.

In this paper, we will apply the DMM scheme to the anomaly-free 331 model for a single family because it is closer to the SM approach. 331 model being among the subgroups of $E_6$~\cite{Mahanthappa:1990pp,Gursey:1975ki,Ramond:1980uc} grand unification theory is another advantage of this model. One of the most important features of the $E_6$ inspired 331 model is the prediction of 3 different Higgs fields. This means that the up quark and the down quark interact with different Higgs fields. The last Higgs contributes to the mass of the heavy isosinglet quarks. With the DMM scheme it will shown that the quark masses and the CKM mixing angles of the SM can be obtained naturally in agreement with the most recent experimental data.

The extension of the $SU(3)_L\otimes U(1)_X$ flavor group with possible fermion and Higgs-boson representations has been investigated~\cite{Albright:1974nd}. Some have studied these extensions as indistinguishable duplicates of a family as in SM~\cite{Mahanthappa:1991pw,Mahanthappa:1990pp,Mahanthappa:1990xt,Singer:1978zd,Barnett:1977rn,Horn:1977hg,Langacker:1977sd,Lee:1977qs,Yoshimura:1976ex,Segre:1976rc,Ramond:1976jg,Gursey:1976dn,Fritzsch:1976dq,Fayet:1974fj}, while others have looked at them as a multi-family construct~\cite{GomezDumm:1997br,Ozer:1995xi,Ozer:1996jc,Epele:1995vv,Ng:1992st,Pleitez:1993gc,Montero:1992jk,Pleitez:1992xh,Foot:1992rh,Pisano:1992bxx} and implying a natural solution to the problem of SM fermion family number~\cite{Long:1995ctv,Long:1996rfd,Pleitez:1994pu,Long:2016lmj,Singer:1980sw}. Many models lead to gauge anomalies~\cite{Fayet:1974fj,kiyan:2002qw,Kiyan:2002jh}, flavor-changing neutral currents~\cite{Mahanthappa:1990pp,Mahanthappa:1990xt,Singer:1978zd,Horn:1977hg,Langacker:1977sd,Yoshimura:1976ex,Segre:1976rc,Fritzsch:1976dq}, right-handed currents at low scales~\cite{Fritzsch:1976dq}, violation of quark-lepton universality~\cite{Mahanthappa:1991pw,Yoshimura:1976ex,Segre:1976rc,Lebbal:2021yrv,Descotes-Genon:2017ptp}, and flavour anomalies~\cite{Addazi:2022frt}. For instance, some models studied in Ref.~\cite{GomezDumm:1997br,Ozer:1996jc,Epele:1995vv,Ng:1992st,Pleitez:1993gc,Montero:1992jk,Pleitez:1992xh,Foot:1992rh,Pisano:1992bxx} lead to physical inconsistencies which rule them out. Meanwhile, these is in agreement with the SM phenomenology, with the 3 quark and 3 lepton families, and the anomaly of the model is eliminated by the addition of quarks carrying exotic electric charges. The model in Ref.~\cite{Long:1995ctv,Long:1996rfd,Pleitez:1994pu,Long:2016lmj,Singer:1980sw} is also three-family and is in agreement with low energy phenomenology, but does not contain exotic electrically charged fermions. In this study, the quark sector of well known $E_6$ inspired $SU(3)_C\otimes SU(3)_L\otimes U(1)_X$ model is considered. It can be renamed in short as Variant-A of 331 model. Details of this variant are given in Ref.~\cite{Ponce:2001jn}. 

In this work, the Variant-A of 331  model is investigated in the light of DMM. The structure of the paper is as follows: Section~\ref{sec:model} introduces 331 model and its variations. The quark content of the Variant-A is presented. Then, definition of the Higgs fields and new gauge bosons of the model, charged and neutral currents are given. DMM approach is applied to the Variant-A of 331 model. In section~\ref{sec:model_param}, new parameterization of the Variant-A is defined. Numerical analysis, obtained correlation plots, are presented in Section~\ref{sec:numerical_analysis}. Section~\ref{sec:results} contains the results of the analysis, more specifically the input parameters and obtained observable variables for the three most relevant and important benchmark points. Section~\ref{sec:discussion} discusses the features and prospects of the obtained results. Finally, Section~\ref{sec:conclusion} concludes the work.
%
\section{331 Model}
\label{sec:model}

As mentioned earlier, $SU(3)_C\otimes SU(3)_L\otimes U(1)_X$ model is one of the minimal extensions of SM. Various sub-models of this model studied earlier~\cite{Ponce:2001jn} contain no exotic electrically charged particles. Two types of these models are variants (A and B)~\cite{Sanchez:2001ua,Martinez:2001mu} that have no triangle anomalies in one family, and the other two are variants (C and D)~\cite{Ozer:1995xi,Singer:1980sw,Pleitez:1994pu,Ozer:1996jc} that have no triangle anomalies in three families. In three-family models, one has different quantum numbers from the other twos. Here the electroweak gauge group is supposed to be $SU(3)_L\otimes U(1)_X \supset SU(2)_L\otimes U(1)_Y$. It is also assumed that left-handed quarks (color triplets) and left-handed leptons (color singlets) transform under the two basic representations of $SU(3)_L$ ($3$ and $3^\ast$). The gauge boson sector is identical in all models, but they may diverge in their quark and lepton contents and scalar sector. In this paper, quark sector of Variant-A of one family model is considered.

\subsection{Quark content of Variant-A}

The quark structure for this model~\cite{Ponce:2001jn} is as following

\begin{align}
\label{eq:q_rep_1}
\begin{matrix}
Q^{\alpha}_{L} = \left( \begin{array}{c}
u_{\alpha}\\
d_{\alpha}\\
D_{\alpha}
\end{array}
 \right)_{L}
& u^{c}_{\alpha L} & d^{c}_{\alpha L} & D^{c}_{\alpha L}, \\
\quad \quad \left\lbrace  3,3,0\right\rbrace  & \left\lbrace  3^{*},1,-\frac{2}{3}\right\rbrace & \left\lbrace  3^{*},1,\frac{1}{3}\right\rbrace & \left\lbrace  3^{*},1,\frac{1}{3}\right\rbrace
\end{matrix}
\end{align}
where $ \alpha =1, 2, 3 $ correspond to the three families. Numbers in parenthesis refer to $\left( SU(3)_{C},\right.$ $SU(3)_{L}$, $\left.U(1)_{X}\right)$ quantum numbers, where $X$ arising in the electric charge generators of the gauge group is defined as
\begin{align}
    \label{eq:q_charge}
    Q = \frac{1}{2} \lambda_{3L} + \frac{1}{2\sqrt{3}} \lambda_{8L} + X I_{3},
\end{align}
where $ \lambda_{iL}$ ($i=1,\dots,8$) are Gell-Mann matrices for $ SU(3)_{L} $ and $ I_{3} $ is 3-dimensional identity matrix.
\subsection{Higgs and New Gauge Bosons}	
Model contains three Higgs fields, which are $(\phi_1^{-}, \phi_1^{0}, \phi_1^{'0})$, $(\phi_2^{-}, \phi_2^{0}, \phi_2^{'0})$ and $(\phi_3^{0}, \phi_3^{+}, \phi_3^{'+})$. Vacuum Expectation Values (VEV) of Higgs fields are thefollowing:
\begin{center}	
\begin{equation}
	\begin{array}{c}
	 \left\langle \phi_{1} \right\rangle = (0,0,M)^{T} ,\\  
	 \left\langle \phi_{2} \right\rangle = (0,\frac{\eta}{\sqrt{2}},0)^{T},\\
	 \left\langle \phi_{3} \right\rangle = (\frac{\eta \prime}{\sqrt{2}},0,0)^{T}. 
	\end{array}
\end{equation}
\end{center}
where $\eta\sim 250$ GeV ($\eta \prime =\eta$ can be taken for simplicity).

In addition, this model has a total of 17 gauge bosons. One of the gauge fields is the gauge boson associated with $U(1)_X$. 8 of them are $SU(3)_C$ associated gauge bosons. The gauge fields of the electroweak sector can be listed as $W^\pm$, $K^\pm$, $K^0$ and $\bar{K}^0$ with mass for charged currents, and $Z$ and $Z^\prime$ for the neutral currents, which are also massive and uncharged. The masses of the new bosons are proportionate to the symmetry breaking scale of the model, in the order of a few TeV. The masses of the gauge bosons of the electroweak sector can be found using the following expressions:

\begin{subequations}
\label{eq:masses_bosons}
\begin{align}
	&m^{2}_{W^{\pm}} = \frac{g^2}{4} (\eta^{2}+\eta^{\prime 2}),\hspace{2.5cm} \\
	&m^{2}_{Z} = \frac{m^{2}_{W^{\pm}}}{C_W^2},\hspace{3.2cm}\\
	&m^{2}_{K^{\pm}} = \frac{g^2}{4} (2 M^{2}+\eta^{\prime 2}),\hspace{2.5cm} \\
	&m^{2}_{K^{0}(\bar{K}^0)} = \frac{g^2}{4} (2 M^{2}+\eta^{2}),\hspace{2.5cm} \\
	&m^{2}_{Z^\prime} = \frac{g^{2}}{4(3-4S_W^2)}\left[8C_W^2M^2+\frac{\eta^{2}}{C_W^2}+\frac{\eta^{2}(1-2S_W^2)^2}{C_W^2}\right].
\end{align}
\end{subequations}
where $C_W$ and $S_W$ are the cosine and sine of the electroweak mixing angle respectively with experimental value of $S_W^2 = 0.23122$. It should be emphasized that in addition to the SM, there are five new gauge bosons, which may lie within the detection limits of the Large Hadron Collider (LHC), as we assume their masses to be in the order of a few TeV. Limitations on the masses of these particles have been identified by the absence of certain types of expected events~\cite{Zyla:2020zbs}. By using last ATLAS~\cite{201968}  and CMS data~\cite{CMS-PAS-EXO-19-019}, a new and restrictive constraint on the mass of the $Z^\prime$ boson was found to be $M_{Z^\prime}~\text{\textgreater}~5.1$ TeV and $M_{Z^\prime}~\text{\textgreater}~4.6$ TeV at $95 \%$ CL, respectively.

In fact, the common feature of many models obtained by extending the SM is the participation of extra heavy gauge bosons~\cite{Zyla:2020zbs}, the charged ones usually denoted by $W^\prime$. In the LHC, $W^{\prime}$ bosons would be observed through production of fermion or electroweak boson pairs resonantly. The most extensively considered signature contains a high-energy electron or muon and large lost transverse energy. Assuming that these new bosons couple with fermions in the SM, restrictive constraints on the mass of $W^{\prime}$ are obtained as $M_{W^{\prime}}~\text{\textgreater}~6$ TeV at $95 \%$ CL~\cite{PhysRevD.100.052013}. Although this limitation does not directly apply to our model, it gives a sense of the masses of the $K^{\pm}$ and $K^0$ bosons.

Charged currents in this model are as follows
\begin{align}
    \label{eq:lag_cc}
\mathcal{L}_{CC} &= -\frac{g}{\sqrt{2}} \left[ \bar{\nu}_{L}^\alpha \gamma^{\mu}e_{L}^{\alpha}W_{\mu}^{+}+\bar{N}_{L}^\alpha \gamma^{\mu}e_{L}^{\alpha}K_{\mu}^{+}+\bar{\nu}_{L}^\alpha \gamma^{\mu} N_{L}^{\alpha} K_{\mu}^{0} + \bar{u}_{\alpha L}\gamma^{\mu}d_{\alpha L} W_{\mu}^{+} \right. \nonumber \\
 & \left. + \Bar{u}_{\alpha L} \gamma^{\mu} D_{\alpha L}  K_{\mu}^{+} - \Bar{D}_{\alpha L} \gamma^{\mu} d_{\alpha L}  K_{\mu}^{0} + \text{h.c.} \right],
\end{align}
and neutral currents are given by
\begin{align}
\label{eq:lag_nc}
    \mathcal{L}^{NC} &= -\frac{g}{2CW} \sum_{f} \left[ \Bar{f} \gamma^{\mu} \left( g_{V}^{\prime} + g_{A}^{\prime} \gamma^5 \right) f Z_{\mu}^{\prime} \right],
\end{align}
where $f$ represents leptons and quarks; $g$, $g_{V}^{\prime}$, and $g_{A}^{\prime}$ are the coupling constants of $SU(3)_L$.

As can be seen from the above expressions, $K^+$ and $K^-$ gauge bosons provide transitions between up sector quarks and new isosinglet D quarks, while $K^0$ and $\bar{K}^0$ gauge bosons mediate the interactions of SM down sector quarks and new isosinglet quarks. 

\subsection{Democratic Approach to the Quark Sector of 331 Model}
\label{sec:dem_quark_331}
The democratic mass matrix (DMM) approach was developed by H. Harari and H. Fritzsch \cite{HARARI1978459,FRITZSCH1979189,FRITZSCH1987391,FRITZSCH1990451,FRITZSCH1994290} to solve the mass hierarchy and mixings' problems, but was unsuccessful in predicting top quark's mass. To remedy this, a number of papers were published, in which DMM was applied to four family SM~\cite{Datta,CELIKEL1995257}. Later, the SM type fourth family fermions were excluded by ATLAS and CMS data~\cite{DJOUADI2012310,collaboration2013searches}. As a consequence, if DMM approach is correct, it will be inevitably applied to an extension of the SM. DMM assumes that Yukawa coupling constants should be approximately the same in the weak interaction Lagrangian. When the mass eigenstates are turned on, fermions gain different masses~\cite{Atag,CIFTCI.055001,CIFTCI.053006}. 

When applying DMM approach to the Variant-A of 331 model, two different basis are defined: $ SU(3)_L\otimes U(1)_X $ symmetry basis, labeled with superscript ``$(0)$'' as in $f^{(0)}$ and the mass basis labeled without superscript as in $f$, where $f$ stands for any fermion particle. Before the electroweak spontaneous symmetry breaking, quarks are grouped as following:

\begin{subequations}
\label{eq:eigenstates}
\begin{align}
	&\left(\begin{matrix} u^{(0)} \\ d^{(0)} \\ D^{(0)} \end{matrix}\right)_{L}, \quad \begin{matrix}u^{c(0)}_{L}, & d^{c(0)}_{L}, & D^{c(0)}_{L}\end{matrix}, \\
	&\left(\begin{matrix} c^{(0)} \\ s^{(0)} \\ S^{(0)} \end{matrix}\right)_{L}, \quad \begin{matrix}c^{c(0)}_{L}, & s^{c(0)}_{L}, & S^{c(0)}_{L}\end{matrix}, \\
	&\left(\begin{matrix} t^{(0)} \\ b^{(0)} \\ B^{(0)} \end{matrix}\right)_{L}, \quad \begin{matrix}t^{c(0)}_{L}, & b^{c(0)}_{L}, & B^{c(0)}_{L}\end{matrix}.
\end{align}
\end{subequations}

In one-family case, all bases are equal. The Lagrangian with the quark Yukawa terms for only one-family case can be written as following:
\begin{equation}
\label{eq:lag_A}
    \mathcal{L}^{Q}_{y} = Q^{T}_{L}C (a_{u}\phi_{3}u^{c}_{L}+a_{d}\phi_{2}d^{c}_{L}+a_{D}\phi_{1}D^{c}_{L}+a_{dD}\phi_{2}D^{c}_{L}+a_{Dd}\phi_{1}d^{c}_{L})+h.c.,
\end{equation}
where $ a_{u} $, $ a_{d} $, $ a_{D} $, $ a_{dD} $ and $ a_{Dd} $ are Yukawa couplings in the $ SU(3)_{L}\otimes U(1)_{X} $ basis and $C$ is the charge conjugate operator.

In this case, we obtain a mass term for the up-quark sector:
 \begin{equation}
 m^{0}_{u}=a_{u}\frac{\eta^u}{\sqrt{2}}\quad (\eta^u= \eta^d=\eta\text{ is taken for simplicity}),
 \end{equation}
 and a mass term for the down-quark sector is given as:
  \begin{equation}
  m^{0}_{dD}= \left( \begin{array}{cc}
  a_{d} \eta^d / \sqrt{2} & \varepsilon a_{d} \eta^d / \sqrt{2} \\
 \varepsilon a_{D} \eta^D / \sqrt{2} & a_{D} \eta^D / \sqrt{2}  
  \end{array}
   \right),
\end{equation}
where $\varepsilon$ is chosen very close to one, and $ \varepsilon a_{d} $ corresponds to the $ a_{dD} $ and $ \varepsilon a_{D} $ corresponds to the $ a_{Dd}$.

In order to obtain mass eigenvalues, we need to diagonalize the above mass matrix. This is done in ref.~\cite{Ciftci:2016hbv} to demonstrate that this approach gives correct $ t $ and $ b $ quark masses in one-family case.
  
Now, we can write three-family quark Yukawa Lagrangian in the $SU(3)_{L}\otimes U(1)_{X}$ basis:
\begin{align}
    \label{eq:lag_2}
    \mathcal{L}^{Q}_{y} &= \displaystyle\sum_{i=1}^{3} Q^{i T}_{L}C (a_{u}\phi_{3}u^{c}_{L}+a_{d}\phi_{2}d^{c}_{L}+a_{D}\phi_{1}D^{c}_{L}+\varepsilon a_{d} \phi_{2}D^{c}_{L}+\varepsilon a_{D}\phi_{1}d^{c}_{L}) \nonumber \\
    &+ \displaystyle\sum_{i=1}^{3} Q^{i T}_{L}C (a_{c}\phi_{3}c^{c}_{L}+a_{s}\phi_{2}s^{c}_{L}+a_{S}\phi_{1}S^{c}_{L}+\varepsilon a_{s} \phi_{2}S^{c}_{L}+\varepsilon a_{S}\phi_{1}s^{c}_{L}) \\
    &+ \displaystyle\sum_{i=1}^{3} Q^{i T}_{L}C (a_{t}\phi_{3}t^{c}_{L}+a_{b}\phi_{2}b^{c}_{L}+a_{B}\phi_{1}B^{c}_{L}+\varepsilon a_{b} \phi_{2}B^{c}_{L}+\varepsilon a_{B}\phi_{1}b^{c}_{L}) + \text{h.c.} \nonumber
\end{align}
 
%
\section{Model Parameterization}
\label{sec:model_param}

Democratic form of all quark mass matrices is broken via a small deviation, represented by $\beta$ and $\gamma$ parameters. Furthermore, the form of deviation is the identical for up, down, and heavy down quarks. In spite of this, deviations are parameterized by independent parameters. Mass matrices for up, down, and heavy down type isosinglet quarks are as follows
\begin{subequations}
\label{eq:MassM}
\begin{align}
\label{eq:MassM_u}
& \mathcal{M}^{0}_{u}=\frac{a^u \eta^u}{\sqrt{2}}\left( \begin{array}{ccc}
1+\gamma_{u} & 1  &  1-\frac{9}{2}\gamma_{u}+\beta_{u} \\
1 & 1-2\gamma_{u} & 1+3\gamma_{u}+\beta_{u} \\
 1-\frac{9}{2}\gamma_{u}+\beta_{u} & 1+3\gamma_{u}+\beta_{u} & 1+4\beta_{u} 
\end{array}
 \right), \\
\label{eq:MassM_d}
& \mathcal{M}^{0}_{d}=\frac{a^d \eta^d}{\sqrt{2}}\left( \begin{array}{ccc}
1+\gamma_{d} & 1 &  1-\frac{9}{2}\gamma_{d}+\beta_{d} \\
1 & 1-2\gamma_{d} & 1+3\gamma_{d}+\beta_{d} \\
 1-\frac{9}{2}\gamma_{d}+\beta_{d} & 1+3\gamma_{d}+\beta_{d} & 1+4\beta_{d} 
\end{array}
\right), \\
\label{eq:MassM_D}
& \mathcal{M}^{0}_{D}=\frac{a^D \eta^D}{\sqrt{2}}\left( \begin{array}{ccc}
1+\gamma_{D} & 1  &  1-\frac{9}{2}\gamma_{D}+\beta_{D} \\
1 & 1-2\gamma_{D} & 1+3\gamma_{D}+\beta_{D} \\
 1-\frac{9}{2}\gamma_{D}+\beta_{D} & 1+3\gamma_{D}+\beta_{D} & 1+4\beta_{D} 
\end{array}
\right).
\end{align}
\end{subequations}

However, down sector quarks and isosinglet down type quarks are further mixed according to Eq.~\eqref{eq:lag_2}:

\begin{equation}
\label{eq:MassM_dD}
    \mathcal{M}^{0}_{dD}=\left(
    \begin{array}{cc} 
        \mathcal{M}_{d}^{0} & \varepsilon_{d} \mathcal{M}_{d}^{0} \\
        \varepsilon_{d} \mathcal{M}_{D}^{0} & \mathcal{M}_{D}^{0} 
    \end{array}\right).
\end{equation}

Masses of down SM and Beyond Standard Model (BSM) isosinglet quarks are obtained by diagonalizing $\mathcal{M}^{0}_{dD}$ 6 by 6 mass matrix.  This mass matrix can be diagonalized via a 6 by 6 unitary matrix $U_{dD}$.  Whereas masses of up sector quarks are obtained by diagonalizing $\mathcal{M}^{0}_{u}$ mass matrix with a 3 by 3 unitary matrix $U_{u}$. Analogous 3 by 3 mixing matrices, $U_{d}$ and $U_{D}$, for down type SM and heavy BSM quarks are defined as unitary matrices that diagonalize d and D blocks of the $\mathcal{M}^{0}_{dD}$ given in Eq.~\eqref{eq:MassM_dD}, respectively. For the sake of simplicity, the phases are considered as zero hereafter. Therefore, diagonalizing matrices will be real orthogonal matrices.

$V^W_{CKM}$, $V^{K^{\pm}}$ and $V^{K^0}$ mixing matrices correspond to the W SM electroweak gauge boson, while $K^{\pm}$ and $K^0$ are BSM heavy gauge bosons, respectively. These mixing matrices are defined via a combinations of 3 by 3 diagonalizing matrices $U_{u}$, $U_{d}$, and $U_{D}$, mentioned earlier, and are given as

\begin{subequations}
\label{eq:CKM}
\begin{align}
& V^W_{CKM}= U_{u} U^{T}_{d} = \left( \begin{array}{ccc}
 V_{ud}  & V_{us}   & V_{ub} \\
 V_{cd}  & V_{cs}   & V_{cb} \\
 V_{td}  & V_{ts}   & V_{tb}
\end{array}
\right), \\
& V^{K^{\pm}} = U_{D} U^{T}_{u} = \left( \begin{array}{ccc}
 V_{Du}  & V_{Dc}   & V_{Dt} \\
 V_{Su}  & V_{Sc}   & V_{St} \\
 V_{Bu}  & V_{Bc}   & V_{Bt}
\end{array}
\right), \\
& V^{K^{0}} = U_{D} U^{T}_{d} = \left( \begin{array}{ccc}
 V_{Dd}  & V_{Ds}   & V_{Db} \\
 V_{Sd}  & V_{Ss}   & V_{Sb} \\
 V_{Bd}  & V_{Bs}   & V_{Bb}
\end{array}
\right).
\end{align}
\end{subequations}

These matrices can be parameterized with three mixing angles and one phase angle:

\begin{equation}
\label{eq:CKMangles}
 V= \left( \begin{array}{ccc}
c_{12}c_{13}  & s_{12}c_{13}   & s_{13}e^{-i\delta} \\
-s_{12}c_{23}-c_{12}s_{23}s_{13}e^{i\delta}  & c_{12}c_{23}-s_{12}s_{23}s_{13}e^{i\delta}   & s_{23}c_{13} \\
s_{12}s_{23}-c_{12}c_{23}s_{13}e^{i\delta}  & -c_{12}s_{23}-s_{12}c_{23}s_{13}e^{i\delta}   & c_{23}c_{13}
\end{array}
\right) 
\end{equation}
where $c_{ij}\equiv cos\theta_{ij}$, $s_{ij}\equiv sin\theta_{ij}$; the angles $\theta_{ij}$ are mixing angles and $\delta$ is CP violating phase angle (its contribution has not been considered in this study). 
%
\section{Numerical analysis}
\label{sec:numerical_analysis}
We perform a numerical scan over all parameter region (7 input parameters, for details see Tab.~\ref{tab:benchmark_param}), first by randomly scanning over large parameter regions, and then by performing a close neighborhood scan over specific regions in order to find a global minimum with higher precision. After making numerical scans, we analyze the correlation between different input parameters, distinctive input parameters and predicted observable variables, as well as between various output observable variables. The correlations presented below will increase the predictive power of the model and assist in probing the model in the current and future phenomenological experiments. Given further are the most striking correlations between these and attempt to explain the origin of correlations for some cases.

In order to obtain the results given below we used this following values for $a$ and $\eta$ (defined in Sec.~\ref{sec:dem_quark_331}) parameters
\begin{subequations}
\label{eq:a_eta_values}
\begin{align}
\label{eq:a_eta_values_ud}
    \frac{a^{u,d} \eta^{u,d}}{\sqrt{2}} &= 56.5~\text{GeV}, \\
\label{eq:a_eta_values_D}
    \frac{a^D \eta^D}{\sqrt{2}} &= 3\times 10^4~\text{GeV}.
\end{align}
\end{subequations}
\begin{figure}[!h]
    \centering
    \begin{subfigure}[t]{0.45\textwidth}
        \includegraphics[width=\textwidth]{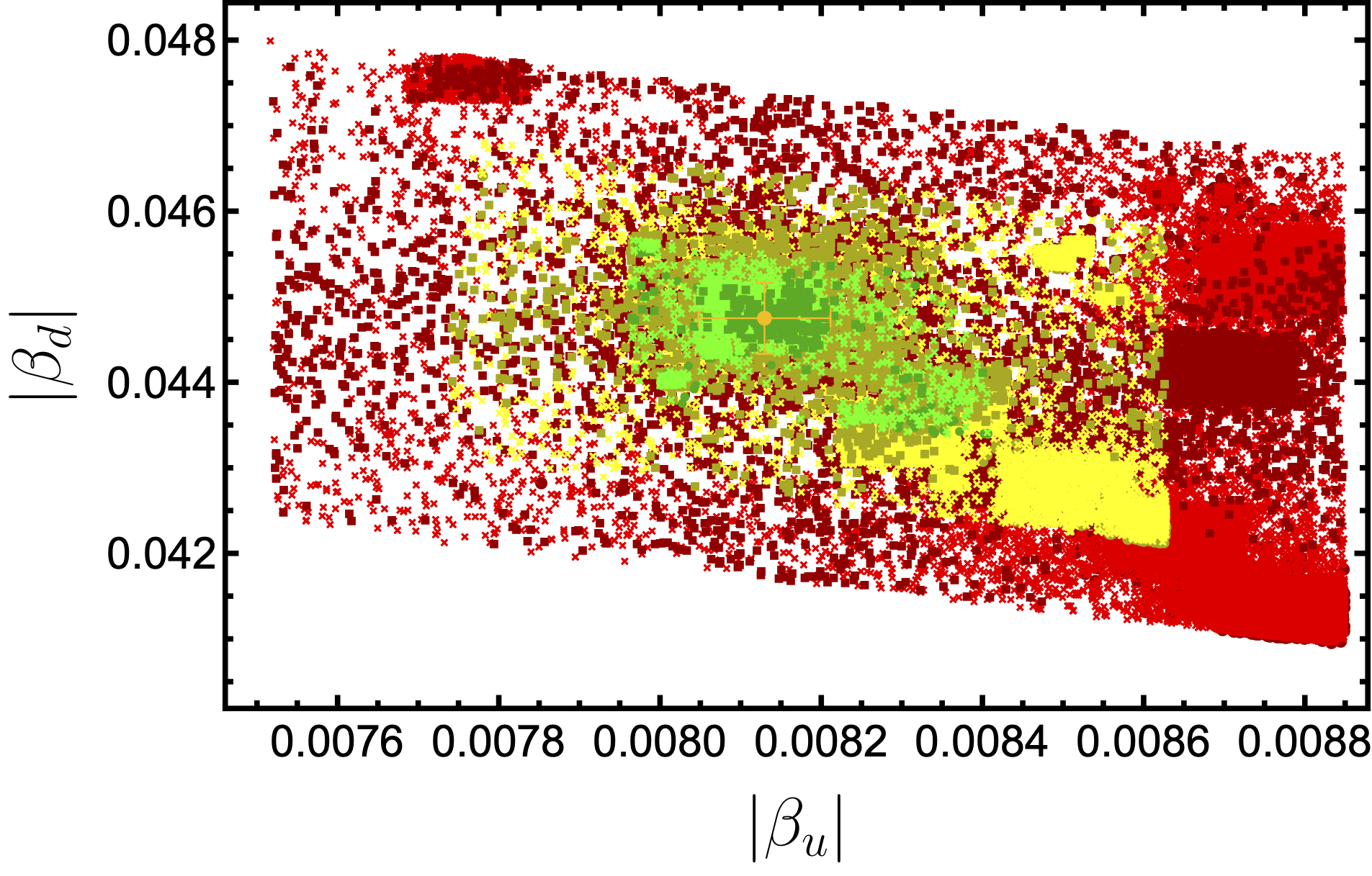}
        \caption{$\beta_u$ vs $\beta_d$ correlation plot.}
        \label{fig:bu_bd_A}
    \end{subfigure}
    \begin{subfigure}[t]{0.45\textwidth}
        \includegraphics[width=\textwidth]{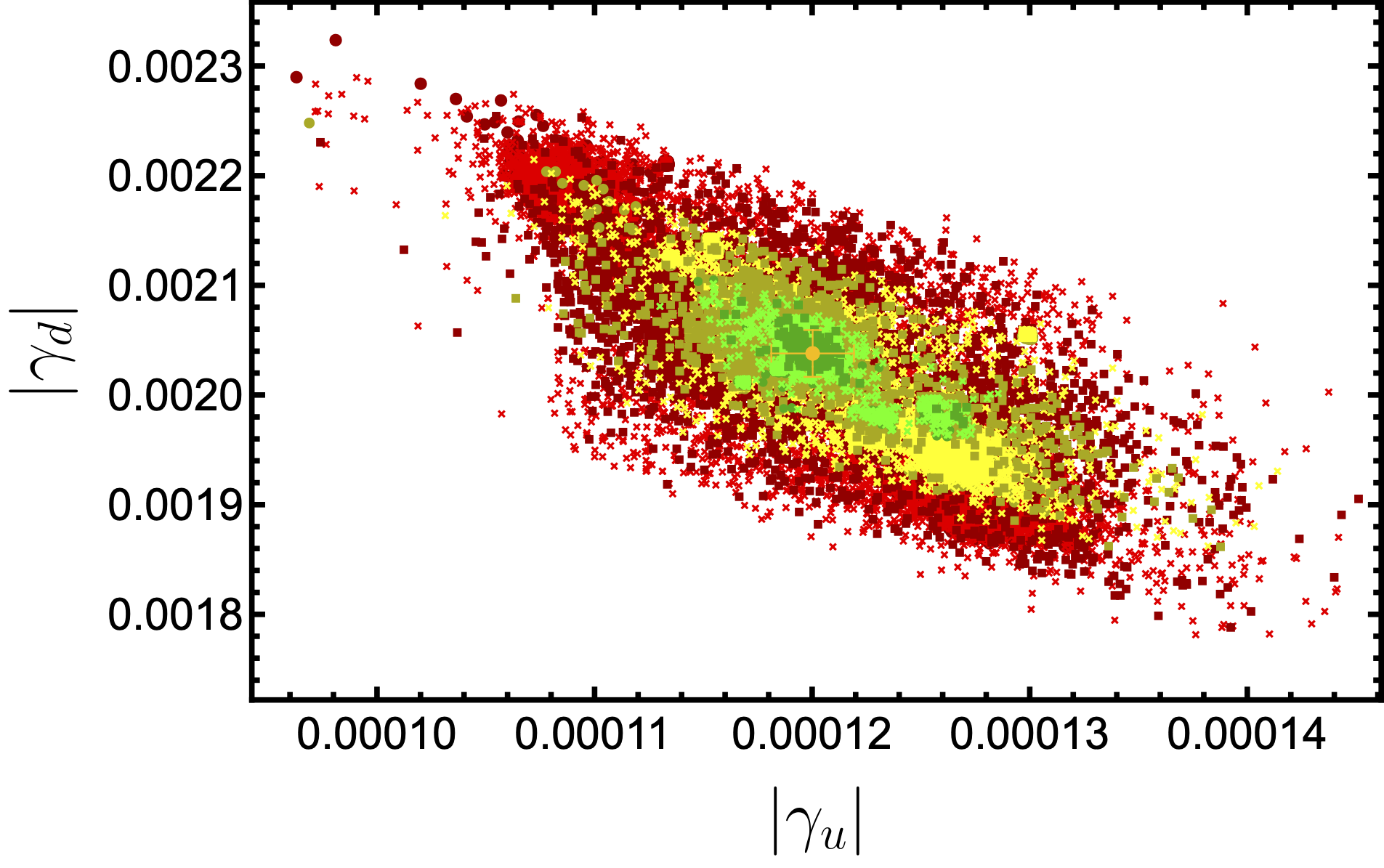}
        \caption{$\gamma_u$ vs $\gamma_d$ correlation plot.}
        \label{fig:gd_gu_A}
    \end{subfigure}
    \begin{subfigure}[t]{0.45\textwidth}
        \includegraphics[width=\textwidth]{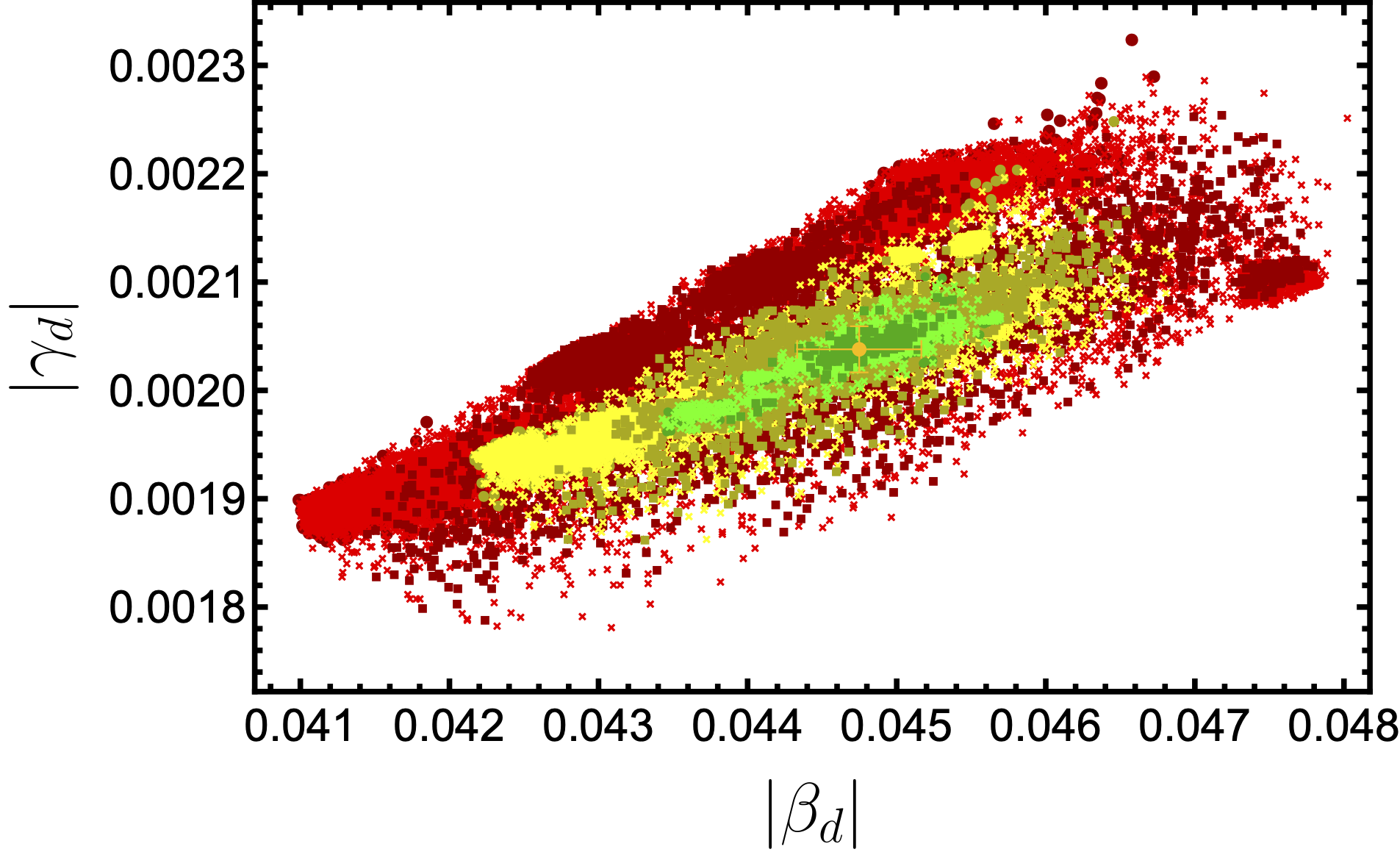}
        \caption{$\beta_d$ vs $\gamma_d$ correlation plot.}
        \label{fig:bu_gu_A}
    \end{subfigure}
    \caption{Selected input correlation plots. Colors represent maximum standard deviation from experimental values. Green, yellow, and red colors stand for $\sigma_{\text{max}}<1$, $2$, and $3$, respectively. Whereas, squares, crosses, and discs correspond to $\left\langle\sigma\right\rangle / \sigma_{\text{max}}$: $0-0.4$, $0.4-0.6$, $0.6-1.0$, respectively.}
    \label{fig:bg_corr_A}
\end{figure}

The most apparent connected patterns between diverse input parameters are shown in Fig.~\ref{fig:bg_corr_A}. As one may have noticed from Fig.~\ref{fig:bu_bd_A}, there is the inverse correlation between $\beta_u$ and $\beta_d$ input parameters. Since $\beta_u$ for up sector behaves identically as does $\beta_d$ for down sector, these inverse correlation is originated from the CKM mixing angles. Fig.~\ref{fig:gd_gu_A} demonstrates the correlation between $\gamma_u$ and $\gamma_d$, which exhibits an inverse correlation as well, analogously with the $\beta$'s case. As a final example for the input parameters Fig.~\ref{fig:bu_gu_A} depicts the direct correlation between input parameters that affect the down sector mass eigenvalues. Other plots for different input parameters show weak or no correlated patters, unlike the ones mentioned earlier.

\begin{figure}[!hb]
    \centering
    \begin{subfigure}[t]{0.45\textwidth}
        \includegraphics[width=\textwidth]{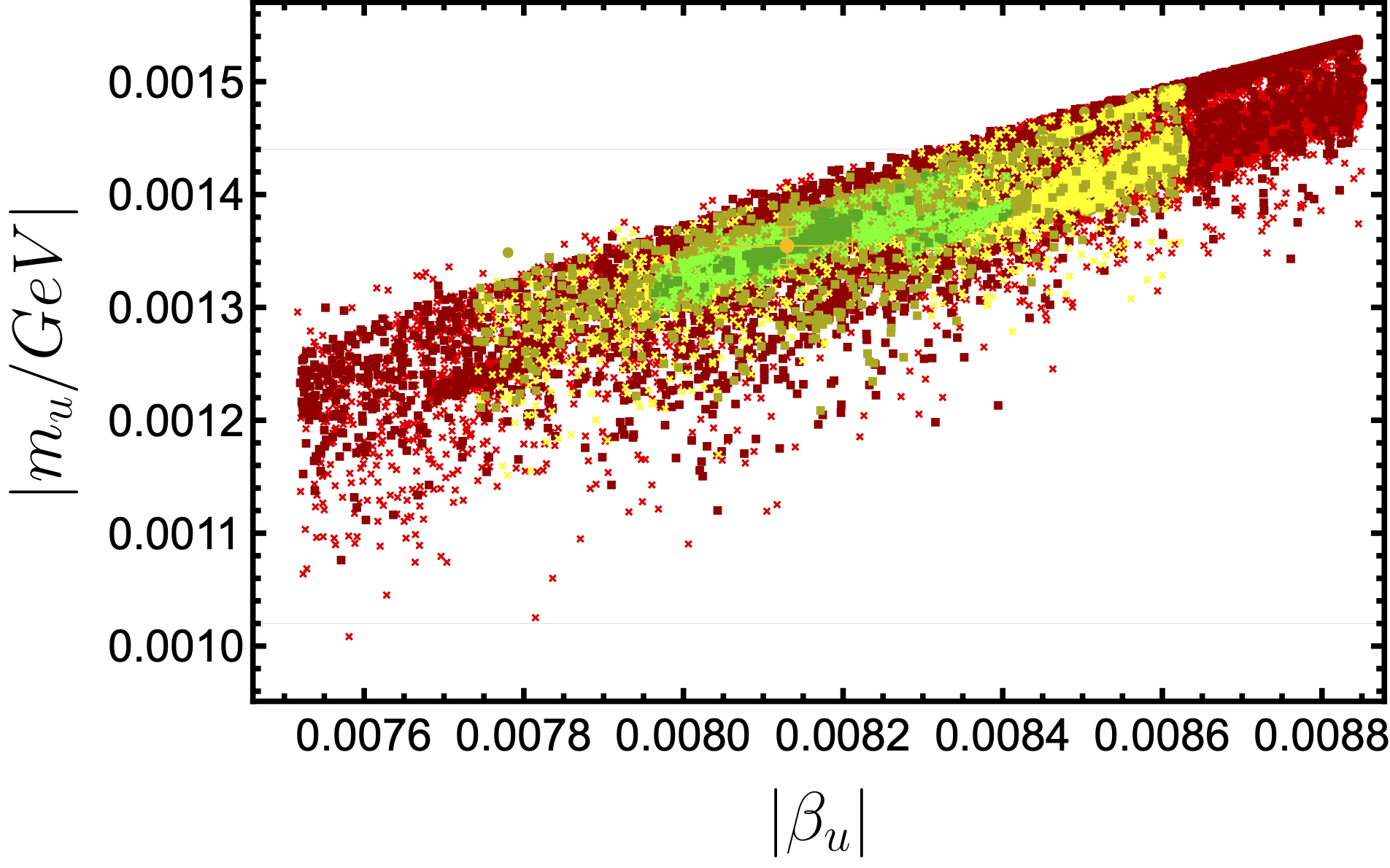}
        \caption{$\beta_u$ vs $m_u$ correlation plot.}
        \label{fig:bu_mu_A}
    \end{subfigure}
     \begin{subfigure}[t]{0.45\textwidth}
        \includegraphics[width=\textwidth]{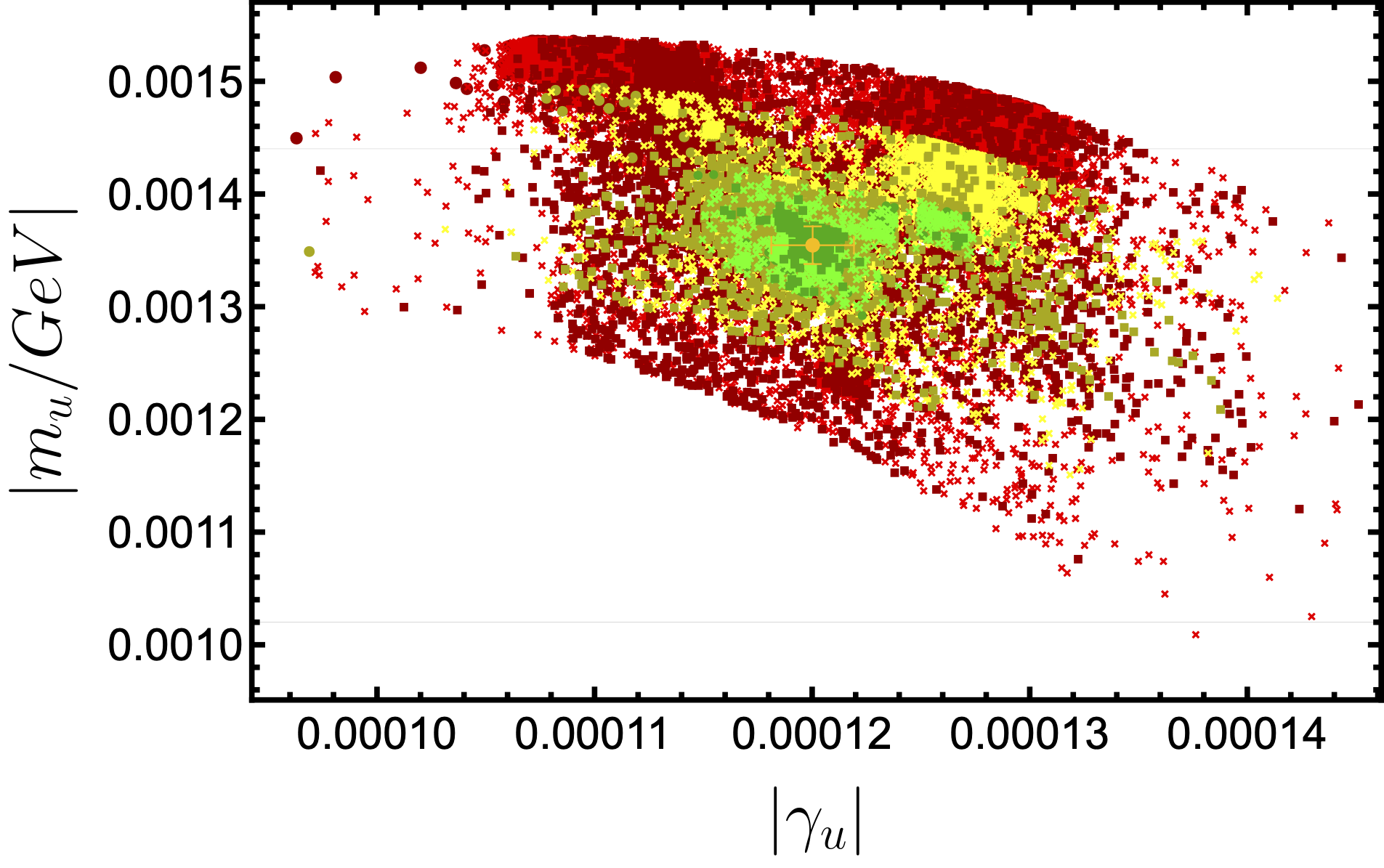}
        \caption{$\gamma_u$ vs $m_u$ correlation plot.}
        \label{fig:gu_mu_A}
    \end{subfigure}
    \begin{subfigure}[t]{0.45\textwidth}
        \includegraphics[width=\textwidth]{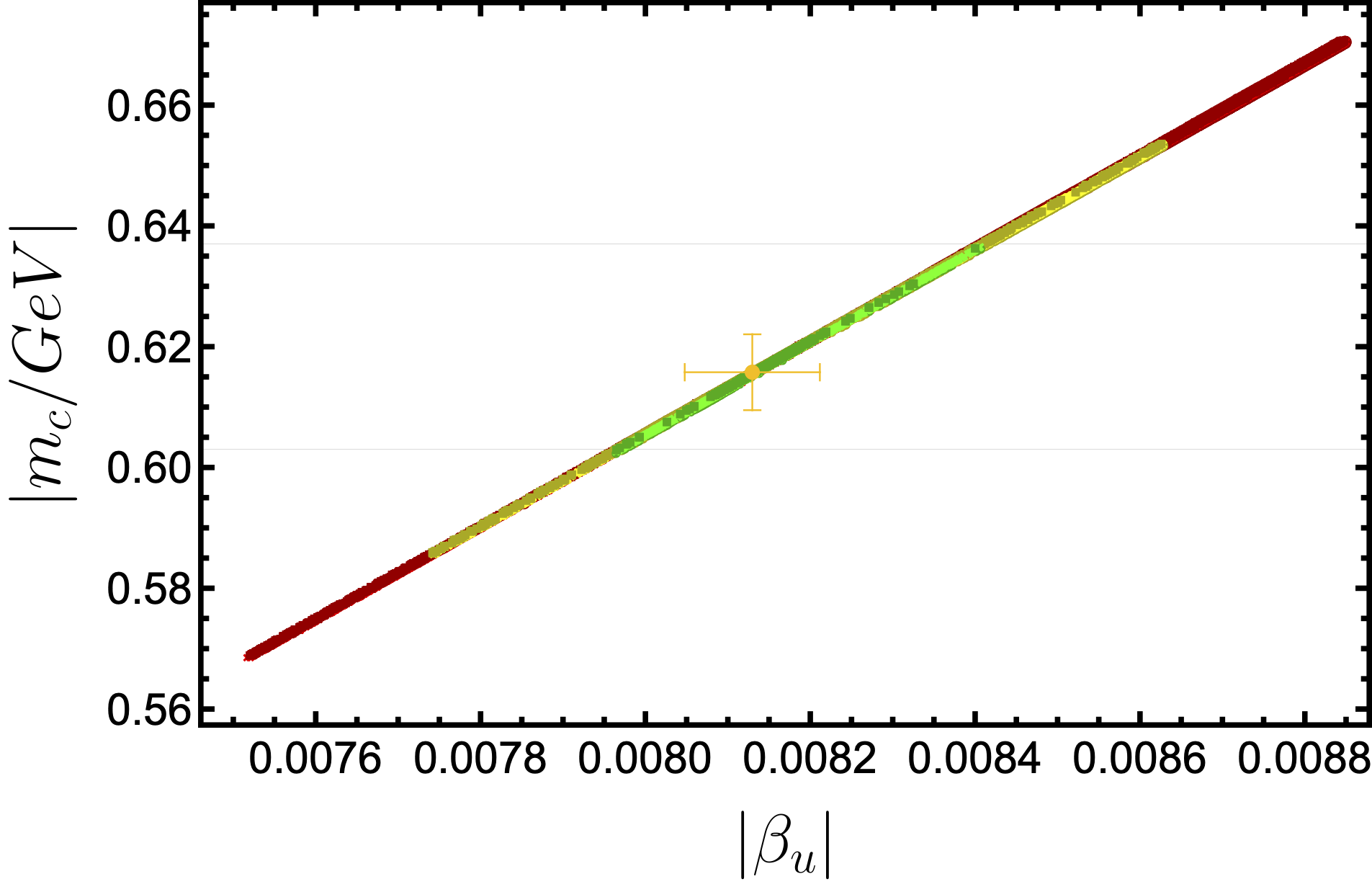}
        \caption{$\beta_u$ vs $m_c$ correlation plot.}
        \label{fig:bu_mc_A}
    \end{subfigure}
    \begin{subfigure}[t]{0.45\textwidth}
        \includegraphics[width=\textwidth]{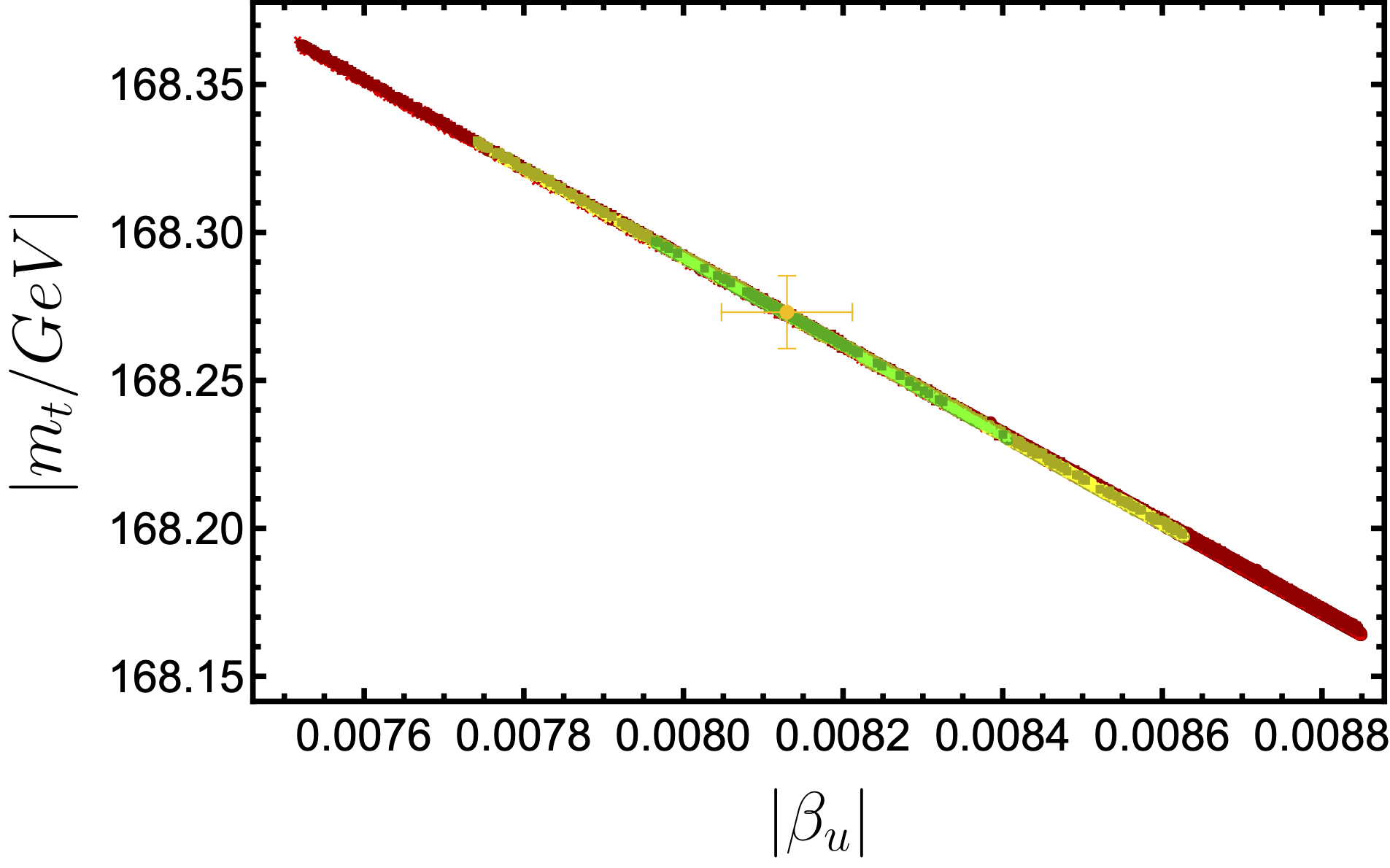}
        \caption{$\beta_u$ vs $m_t$ correlation plot.}
        \label{fig:bu_mt_A}
    \end{subfigure}
\end{figure}
\begin{figure}[!ht]\ContinuedFloat
\centering
    \begin{subfigure}[t]{0.45\textwidth}
        \includegraphics[width=\textwidth]{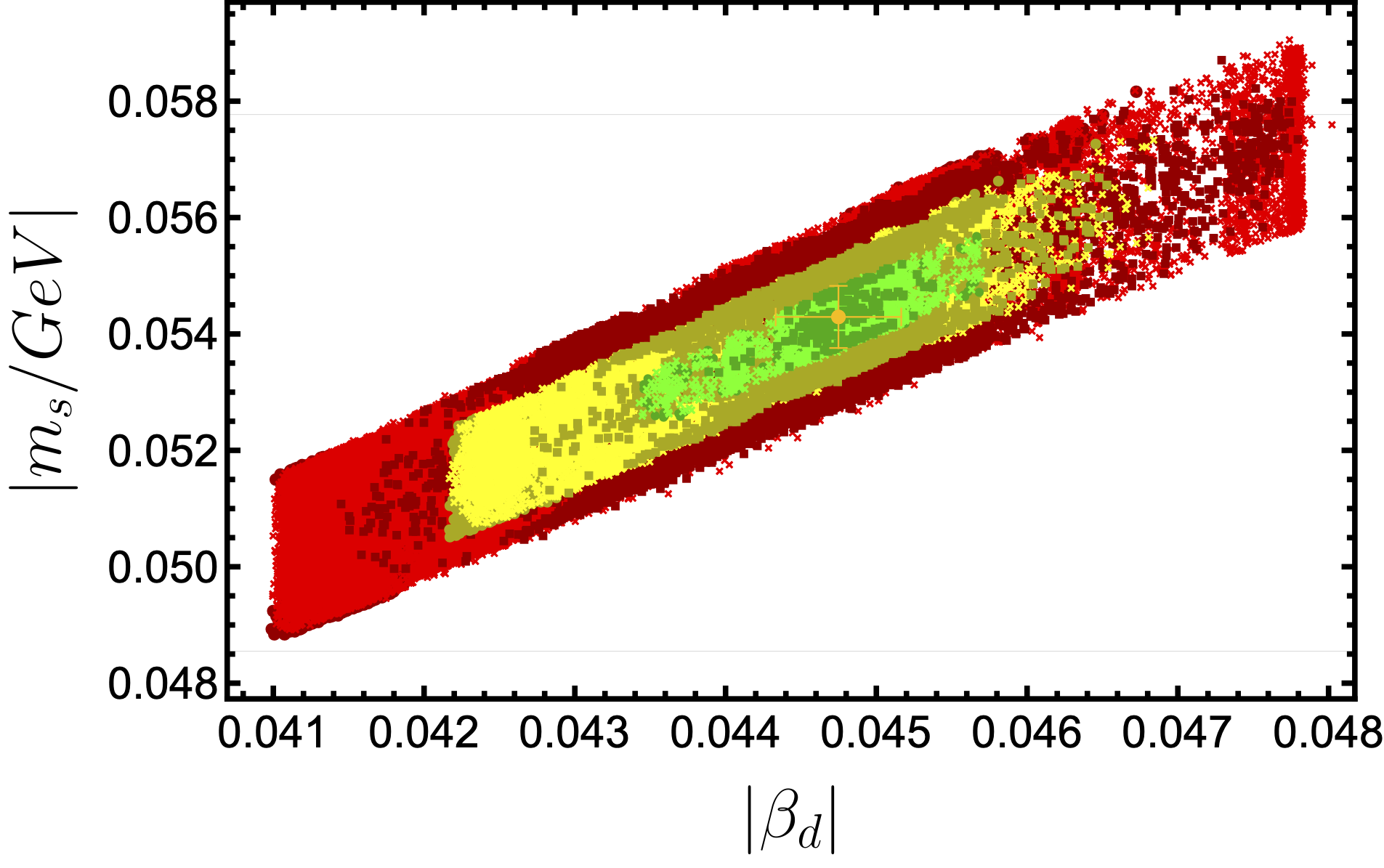}
        \caption{$\beta_d$ vs $m_s$ correlation plot.}
        \label{fig:bd_ms_A}
    \end{subfigure}
    \begin{subfigure}[t]{0.45\textwidth}
        \includegraphics[width=\textwidth]{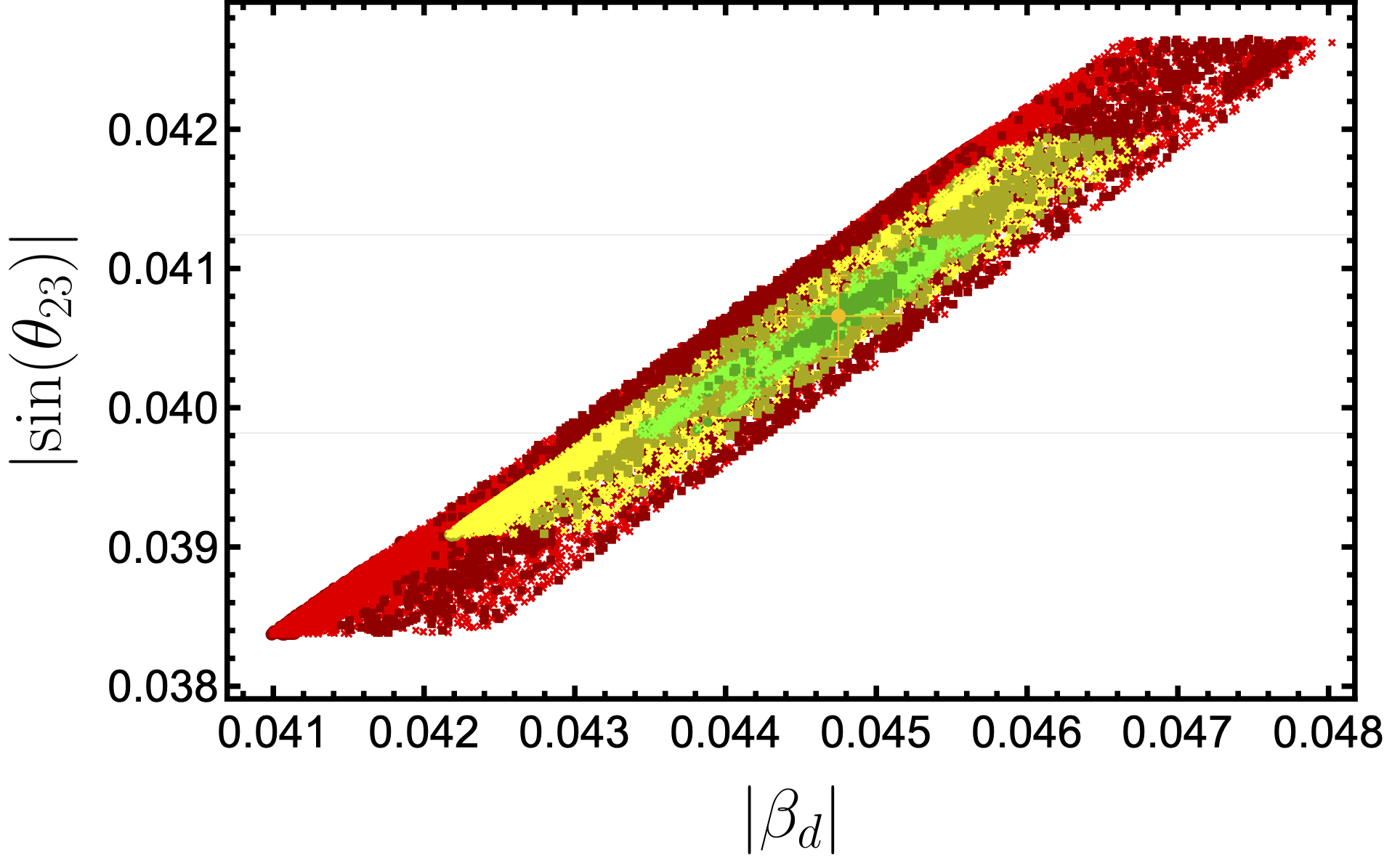}
        \caption{$\beta_d$ vs $\sin\left(\theta_{23}^{\text{\tiny{CKM}}}\right)$ correlation plot.}
        \label{fig:bd_s23_A}
    \end{subfigure}
    \begin{subfigure}[t]{0.45\textwidth}
        \includegraphics[width=\textwidth]{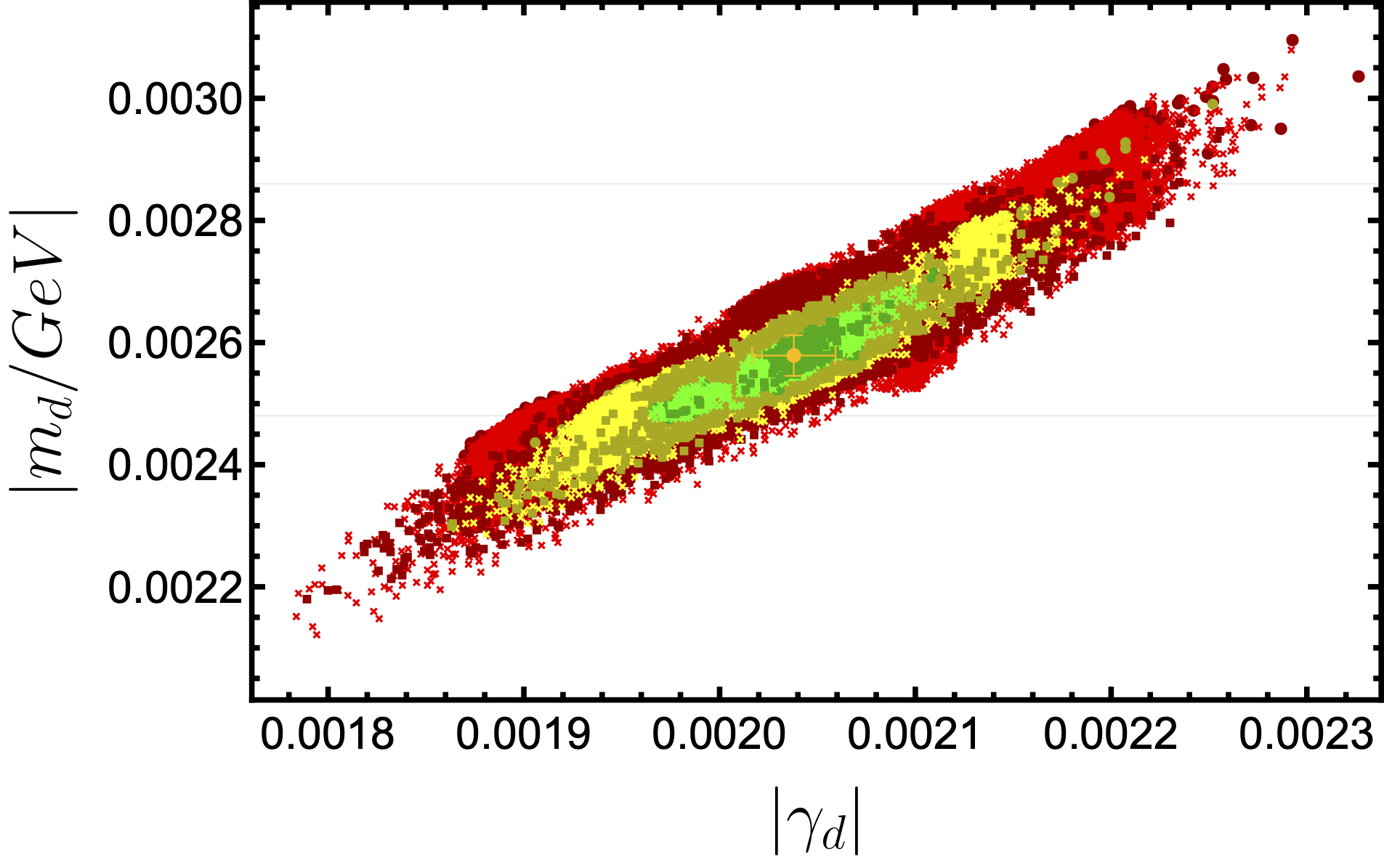}
        \caption{$\gamma_d$ vs $m_d$ correlation plot.}
        \label{fig:gd_md_A}
    \end{subfigure}
    \begin{subfigure}[t]{0.45\textwidth}
        \includegraphics[width=\textwidth]{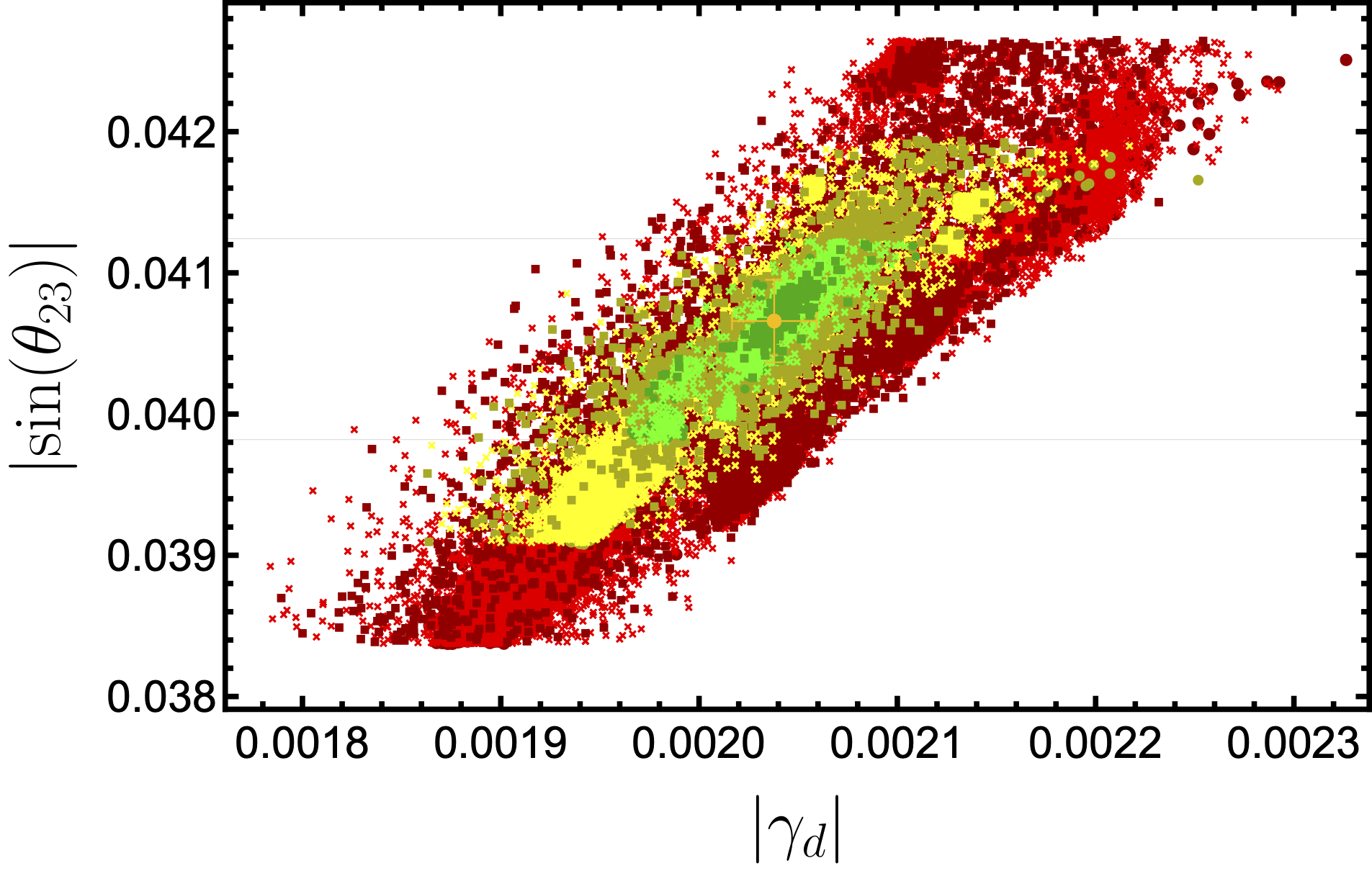}
        \caption{$\gamma_d$ vs $\sin\left(\theta_{23}^{\text{\tiny{CKM}}}\right)$ correlation plot.}
        \label{fig:gd_s23_A}
    \end{subfigure}
    \begin{subfigure}[t]{0.45\textwidth}
        \includegraphics[width=\textwidth]{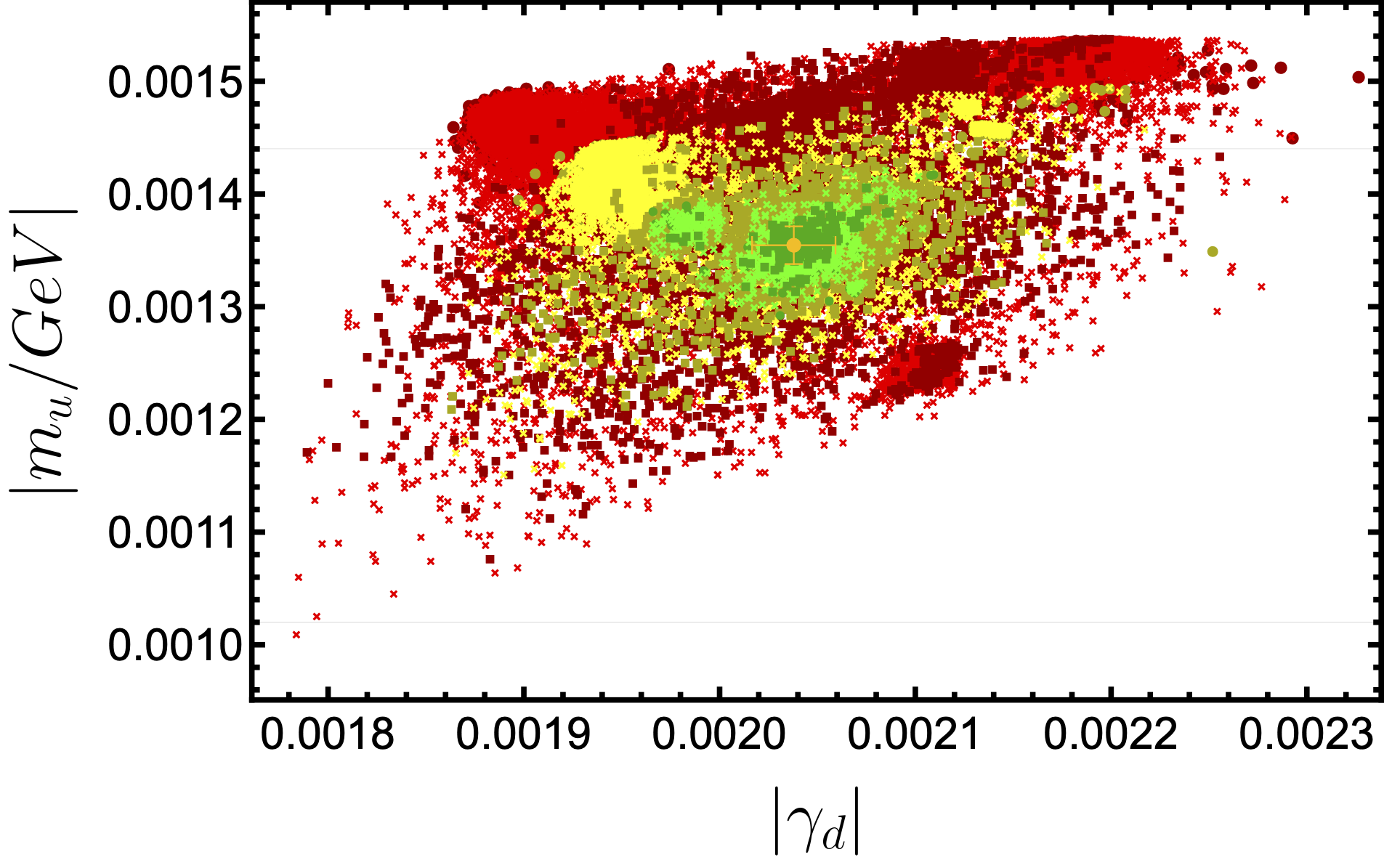}
        \caption{$\gamma_d$ vs $m_u$ correlation plot.}
        \label{fig:gd_mu_A}
    \end{subfigure}
    \caption{Selected correlation plots between input parameters and observable variables. Grid lines indicate a one standard deviation regions of the experimental data. Green, yellow, and red colors stand for $\sigma_{\text{max}}<1$, $2$, and $3$, respectively. Whereas, squares, crosses, and discs correspond to $\left\langle\sigma\right\rangle / \sigma_{\text{max}}$: $0-0.4$, $0.4-0.6$, $0.6-1.0$, respectively.}
    \label{fig:bg_observ_corr_A}
\end{figure}

\begin{figure}[!h]
    \centering
    \begin{subfigure}[t]{0.45\textwidth}
        \includegraphics[width=\textwidth]{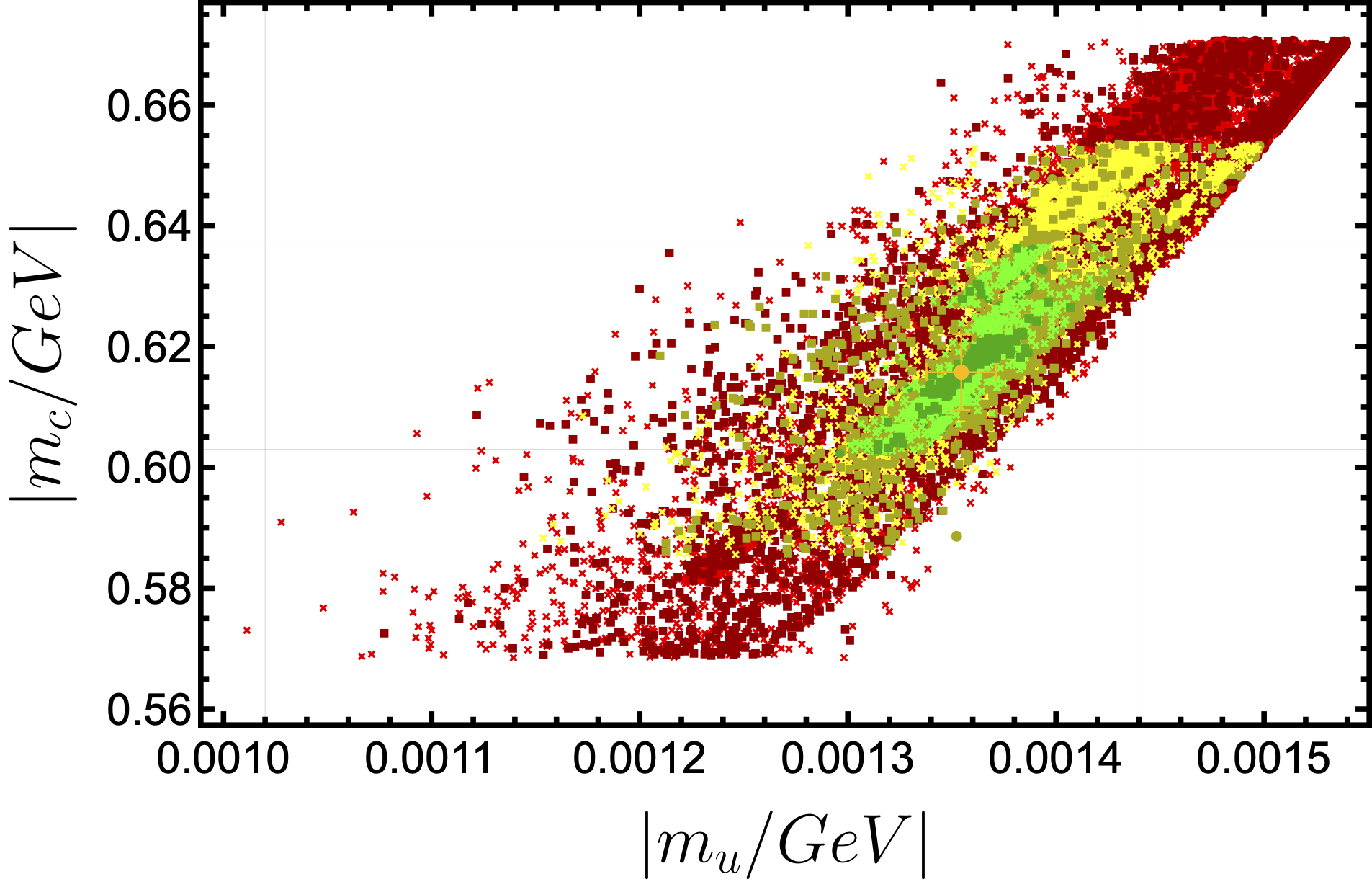}
        \caption{$m_u$ vs $m_c$ global fit distribution graph.}
        \label{fig:mu_mc_A}
    \end{subfigure}
    \begin{subfigure}[t]{0.45\textwidth}
        \includegraphics[width=\textwidth]{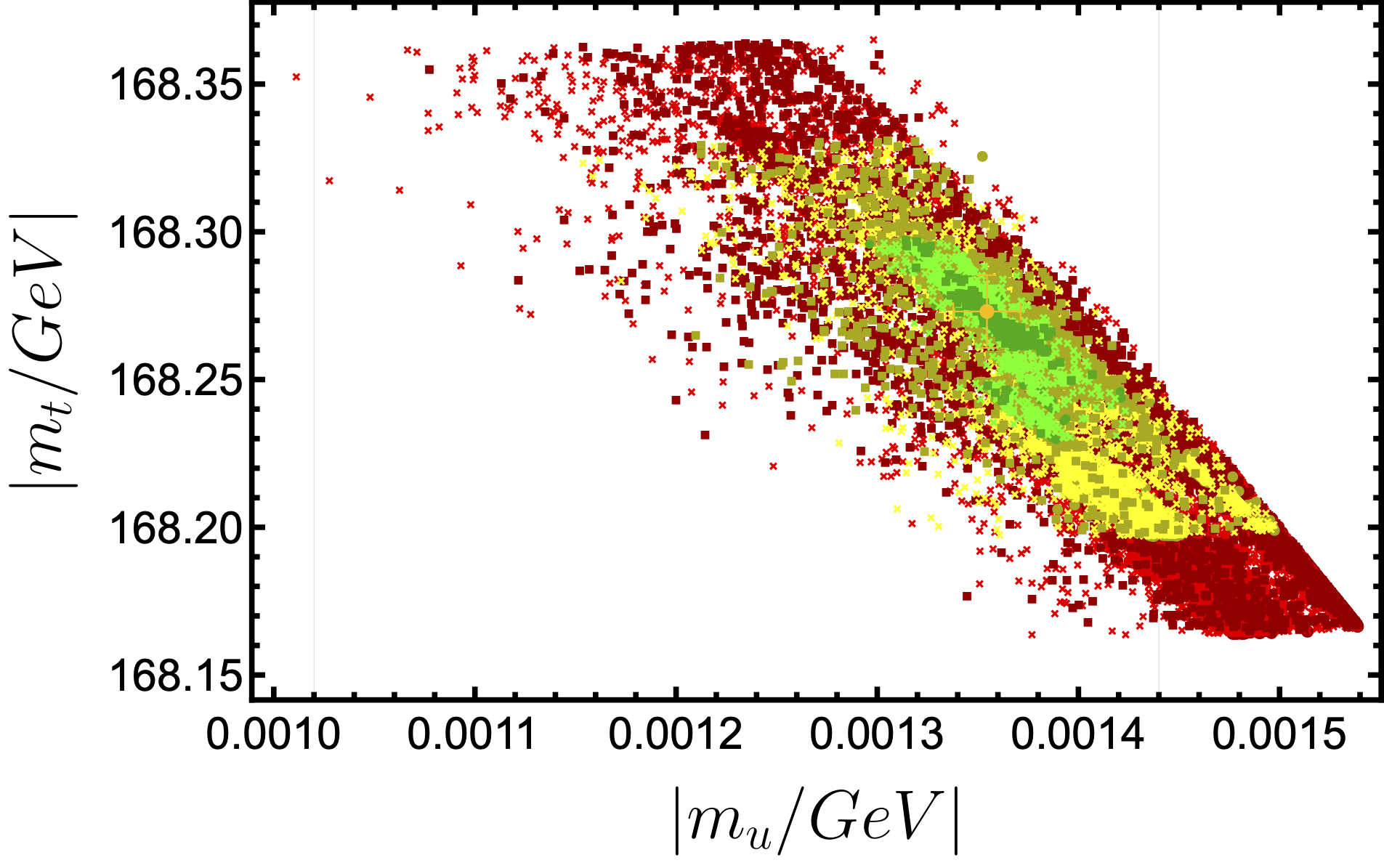}
        \caption{$m_u$ vs $m_t$ global fit distribution graph.}
        \label{fig:mu_mt_A}
    \end{subfigure}
    \begin{subfigure}[t]{0.45\textwidth}
        \includegraphics[width=\textwidth]{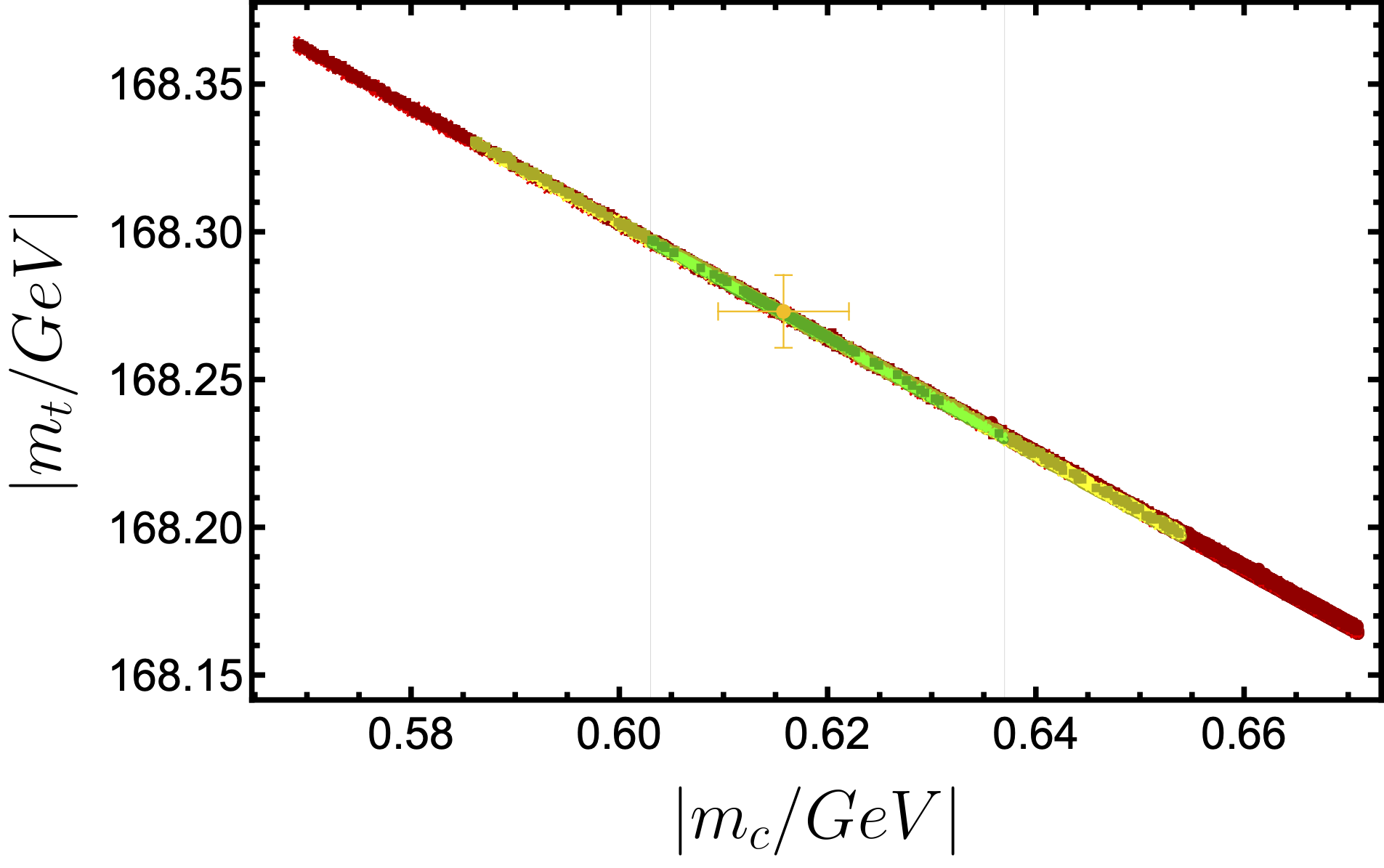}
        \caption{$m_c$ vs $m_t$ global fit distribution graph.}
        \label{fig:mc_mt_A}
    \end{subfigure}
    \begin{subfigure}[t]{0.45\textwidth}
        \includegraphics[width=\textwidth]{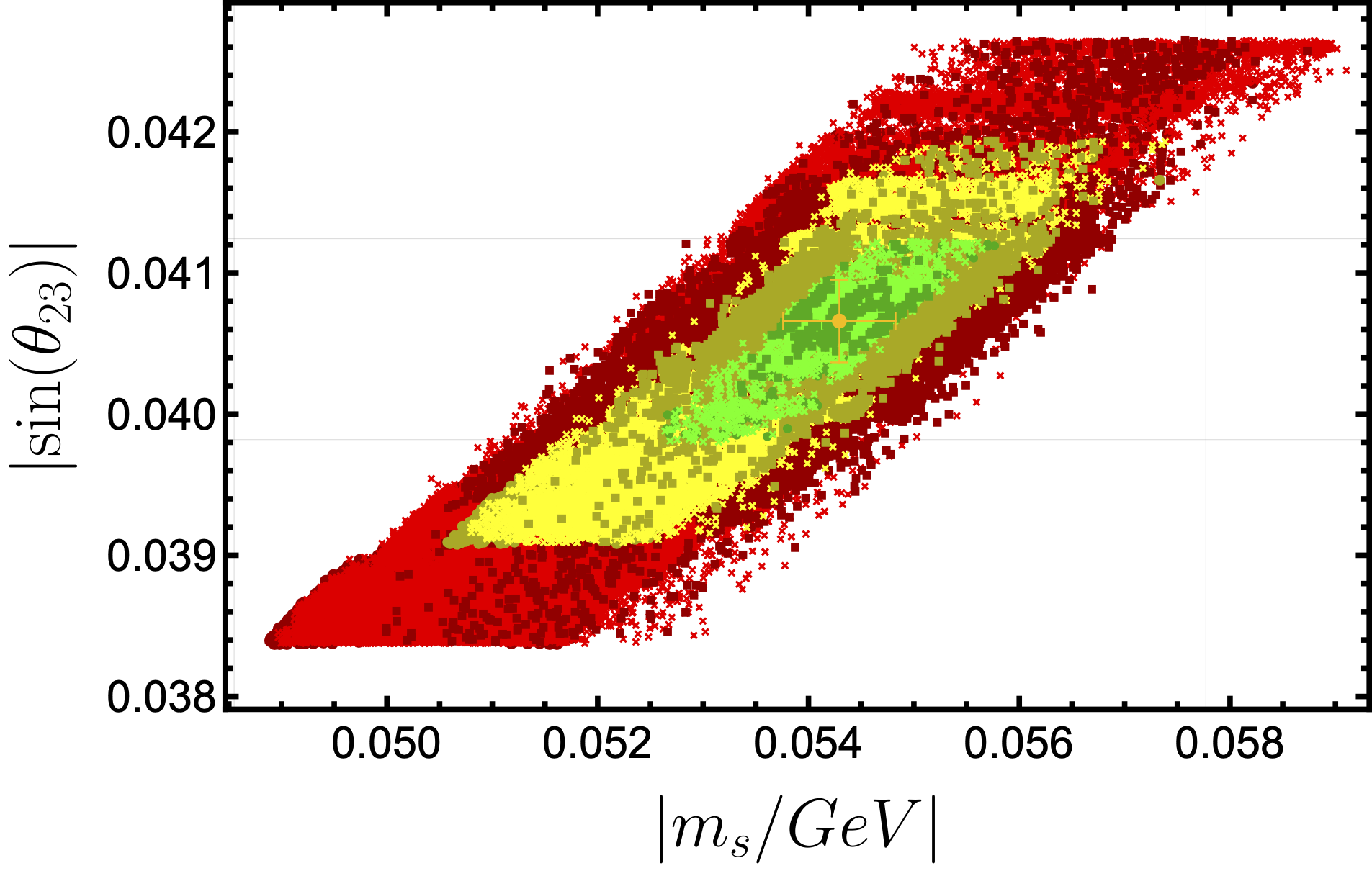}
        \caption{$m_s$ vs $\sin \left(\theta_{23}^{\text{\tiny{CKM}}}\right)$ global fit distribution graph.}
        \label{fig:ms_s23_A}
    \end{subfigure}
    \caption{Selected observable correlation plots. Grid lines indicate a one standard deviation regions of the experimental data. Green, yellow, and red colors stand for $\sigma_{\text{max}}<1$, $2$, and $3$, respectively. Whereas, squares, crosses, and discs correspond to $\left\langle\sigma\right\rangle / \sigma_{\text{max}}$: $0-0.4$, $0.4-0.6$, $0.6-1.0$, respectively.}
    \label{fig:observ_corr_A}
\end{figure}

The plots in the Fig.~\ref{fig:bg_observ_corr_A} demonstrate important dependence of some observable variables on model input parameters. For instance, plot in Fig.~\ref{fig:bu_mu_A} shows the direct dependence of $m_u$, lightest eigenvalue of the up quark sector, on its most sensitive input parameter, $\beta_u$. On the other hand, from Fig.~\ref{fig:gu_mu_A} one can see that there is a inverse non-linear dependence of $m_u$ on $\gamma_u$. Strong correlations are observed between $\beta_u$ and $m_c$, as well as between $\beta_u$ and $m_t$, direct linear and inverse linear, respectively for $m_c$ and $m_t$ up sector quark mass eigenvalues. As a consequence, there is strong correlation between $m_c$ and $m_t$~\ref{fig:mc_mt_A} and all up sector eigenvalues show strong dependence on $\beta_u$ input parameter for the model at hand. Regarding the down quark sector, analogous correlation can be seen between $\beta_d$-$m_s$ and $\beta_d$-$\sin\left(\theta_{23}^{\text{\tiny{CKM}}}\right)$ Fig.~\ref{fig:bd_ms_A} and Fig.~\ref{fig:bd_s23_A}, which exhibit proportional almost linear and direct-linear behaviour, respectively. $\gamma_d$, similar to the situation in the up sector Fig.~\ref{fig:gu_mu_A}, has a strongest influence on the $m_d$, lightest eigenvalue of the down quark sector, Fig.~\ref{fig:gd_md_A}, with a direct behaviour. Furthermore, the direct proportionality between $m_d$ and $\gamma_d$ can be seen from Eq.~\eqref{eq:MassM_d}, for which the lightest eigenvalue($m_d$) approaches to zero as $\gamma_d$ goes to zero.

From the above analysis it can be concluded that $\gamma_{u,d}$ has noticeable influence on the lightest eigenvalue of its respective sector, whereas $\beta_u$ effects all up sector quark masses and $\beta_d$ effects $m_s$ and $\sin\left(\theta_{23}^{\text{\tiny{CKM}}}\right)$. This dependence of $m_b$ on $\beta_d$ is absent due to the mixing of SM down quark sector with BSM heavy quarks.

The most striking correlations of the CKM mixing angles are observed for $\sin\left(\theta_{23}^{\text{\tiny{CKM}}}\right)$ mixing angle, which is directly proportional and completely determined by the two input parameters $\beta_d$ and $\gamma_d$ Fig.~\ref{fig:bd_s23_A} and Fig.~\ref{fig:gd_s23_A}. The other mixing angles exhibit more complex correlations with the input parameters.

Plots in the Figs.~\ref{fig:mu_mc_A},~\ref{fig:mu_mt_A}, and~\ref{fig:mc_mt_A} demonstrate a direct, inverse, and inversely linear correlation between all three up sector quark masses, respectively. This is an immediate consequence of the fact that all three strongly depend on the $\beta_u$ input parameter, Figs.~\ref{fig:bu_mu_A},~\ref{fig:bu_mc_A}, and ~\ref{fig:bu_mt_A}.

Taking the $\gamma_u \ll \beta_u \ll 1$ limit in Eq.~\eqref{eq:MassM_u}, we obtain $m_u\propto 2\gamma_u \ll m_c,m_t$ and $m_t/m_c \approx 5.125 + 7.11806 \beta_u + 2.25 \beta_u^{-1}$, which corresponds to the behaviour $m_t \propto -m_c + \emph{const.}$ (Fig.~\ref{fig:mc_mt_A}). 

Fig.~\ref{fig:ms_s23_A} shows a correlation between down quark sector mass, $m_s$, and CKM mixing angle, $\sin \left(\theta_{23}^{\text{\tiny{CKM}}}\right)$. This can be seen from the direct linear dependence of $\sin \left(\theta_{23}^{\text{\tiny{CKM}}}\right)$ on $\beta_d$ and $\gamma_d$, Fig.~\ref{fig:bd_s23_A} and Fig.~\ref{fig:gd_s23_A}, respectively. Similarly, $m_s$ has medium directly proportional dependence on $\beta_d$ (Fig.~\ref{fig:bd_ms_A}), as well as, weaker dependence on $\gamma_d$.

%
During the numerical analysis, in some cases the obtained output observable variables may have been negative. In which case, the negative sign has been dropped, in the view of the phases related argument made in the paragraph after the Eq.~\eqref{eq:MassM_dD}.
%
\section{Results}
\label{sec:results}
In this section the results of the model predictions are presented and elaborated on. This, $E_6$ motivated, variation of 331 model predicts up and down quark masses, as well as, CKM mixing angles for total of seven input parameters. Up, down, and down type isosinglet quarks are controlled by two parameters each, and one mixing parameter, $\varepsilon$, between light and heavy down quarks. The input parameters for the three most relevant and important benchmark points are collected in Tab.~\ref{tab:benchmark_param}. The first benchmark point (BP1) was obtained as a point with the smallest $\chi^2$ of approximately $0.777$, which has maximum deviation from experimental results of $0.586~\sigma$, refer to Eq.~\eqref{eq:sigma_def} for details. On the other hand, the second benchmark point (BP2) is defined as the point of a parameter scan with the smallest set of deviations for all nine observable variables at hand with a maximum deviation of $\sim0.501~\sigma$. As last, we give an average point for all data points obtained with $\forall \sigma_{\text{max}}\leq1$ as a third benchmark point (BP3), labeled as BP3$_{\langle \rangle}$ in Tab.~\ref{tab:benchmark_obs}, whereas the spread (error) of all points with $\forall \sigma_{\text{max}}\leq1$ is indicated as \emph{Spread}. The $\sigma$ is defined as follows

\begin{align}
    \label{eq:sigma_def}
    \sigma &= \left|\frac{x_{\text{exp}}-x_{\text{th}}}{x_{\text{err}}}\right|,
\end{align}
where $x$ represents any of the observable variables from Tab.~\ref{tab:benchmark_obs}, \emph{exp.} stands for the experimentally obtained value, \emph{th} corresponds to the simulated value from scan run, and lastly, \emph{err.} means the error for the experimentally obtained value.

\begin{figure}[!h]
    \centering
    \includegraphics[width=0.7\textwidth]{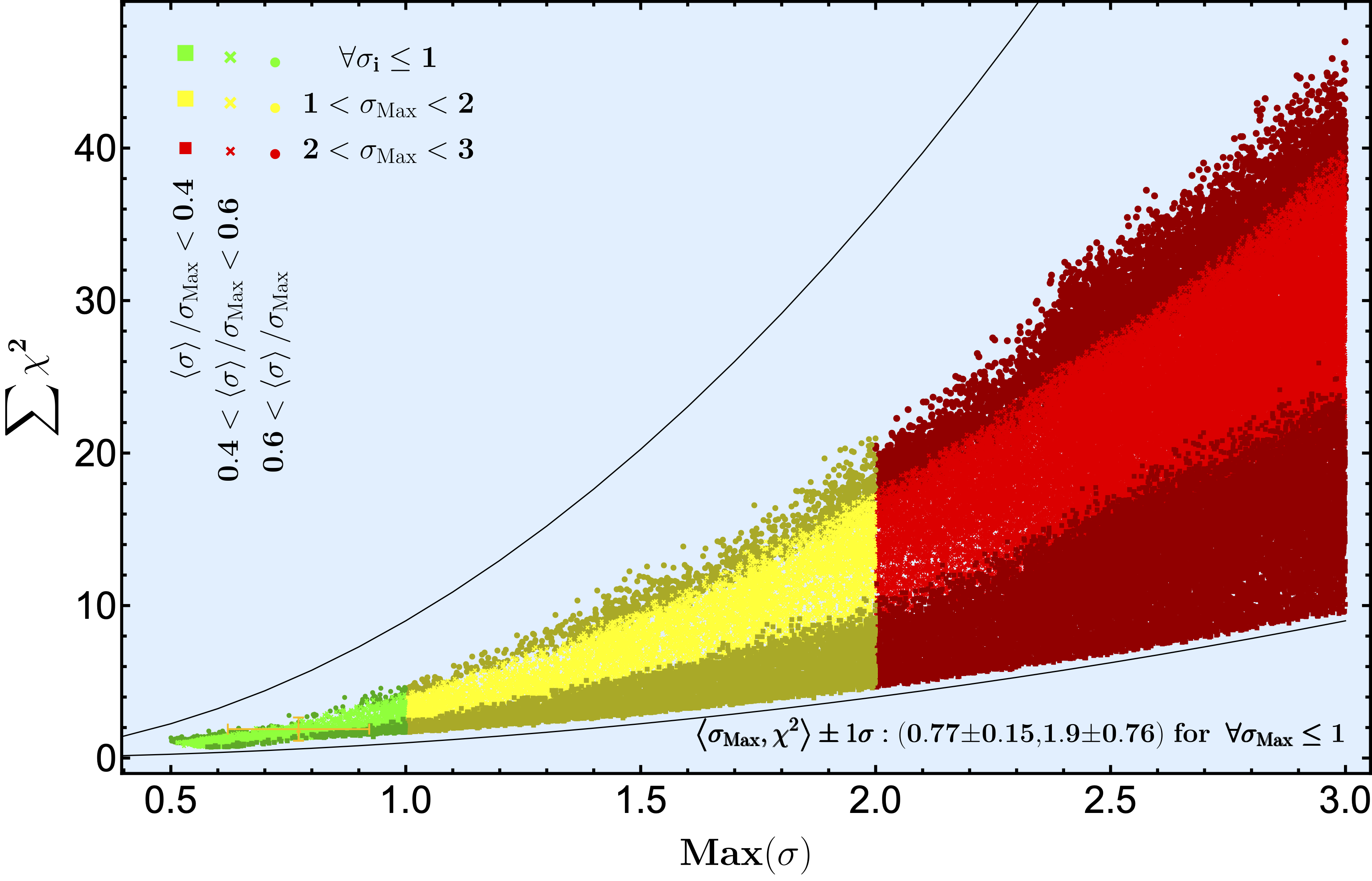}
    \caption{Model global fit vs maximum deviation (up to $3\sigma$) distribution graph. The solid curves stand for upper and lower theoretical limits and $\langle\rangle$ stand for the mean value.}
    \label{fig:max_sigma_chi2}
\end{figure}

\begin{table}[!h]
    \centering
    {\footnotesize
    \begin{tabular}{cllll}
        \hline\hline
        \text{par.} & $\text{BP1}$ & $\text{BP2}$ & $\text{BP3}_{\langle \rangle}$ & $\text{BP3}_{\text{spread}}$ \\ \hline
        $\beta_u$ & -0.008127116503406678 & -0.008074064951393775 & -0.0081294 & 0.0000816256 \\
        $\gamma_u$ & 0.00012023151170229749 & 0.00012095573963901077 & 0.000120017 & 0.0000018874 \\
        $\beta_d$ & 0.044756546016957506 & 0.04508573236580676 & 0.0447492 & 0.000416564 \\
        $\gamma_d$ & 0.002037220066695815 & 0.0020439360347433004 & 0.00203794 & 0.0000212662 \\
        $\beta_D$ & 0.04796738067655066 & 0.04787190321609066 & -0.0474458 & 0.000806016 \\
        $\gamma_D$ & -0.05988836055185294 & -0.06057801433337951 & 0.0605905 & 0.00131765 \\
        $\varepsilon$ & 1.0080400994427337 & 1.0080578648812581 & 1.00563 & 0.00575238 \\
        \hline\hline
    \end{tabular}
    }
    \caption{Model input parameters for the several benchmark points given in Tab.~\ref{tab:benchmark_obs}}
    \label{tab:benchmark_param}
\end{table}
The parameter scan is very sensitive to the input parameters' value, therefore we keep up to twenty decimal places. For a total of seven input parameters the best result for $\chi^2$ that was obtained is given in the 4'th and 5'th columns of Tab.~\ref{tab:benchmark_obs} with $\chi^2\approx 0.777$. As can be seen, the largest contribution to the $\chi^2$ comes from $m_u$ and $m_d$, whereas the second and third generation quark masses of up and down sectors contribute a much smaller error to the $\chi^2$. Next, as a result of the search for the smallest combination of the deviations from the experimental values (2'nd and 3'rd columns of Tab.~\ref{tab:benchmark_obs}), the best point achieved is given in the 6'th and 7'th columns of Tab.~\ref{tab:benchmark_obs} with $\chi^2\approx 1.491$ and $\sigma_{\text{max}}\approx 0.501$. Lastly, we collect all points with maximum deviations ($\sigma_{\text{max}}\leq1$) to generate mean and spread values, given in the 8'th and 9'th columns of Tab.~\ref{tab:benchmark_obs} with $\chi^2\approx 1.895$. These values show the location and the size of the area with the deviations from experimental values less than one (green area in Fig.~\ref{fig:max_sigma_chi2}).
\begin{table}[!h]
    \centering
    {\footnotesize
       \begin{tabular}{ccccccccc}
        \hline\hline
        \scriptsize{\text{Observable}} & \multicolumn{2}{c}{\text{Experimental}} & \multicolumn{2}{c}{\text{BP1}} & \multicolumn{2}{c}{\text{BP2}} & \multicolumn{2}{c}{\text{BP3}} \\ \hline
        & \text{Value} & \text{Err.} & \text{Value} & $\sigma$ & \text{Value} & $\sigma$ & $\langle \rangle$ & Spread \\ \hline
        $m_u$ \scriptsize{(\text{MeV})} & 1.23 & 0.21 & 1.35 & 0.59 & 1.33 & 0.50 & 1.35 & 0.02 \\
        $m_c$ \scriptsize{(\text{MeV})} & 620 & 17 & 616 & 0.26 & 611 & 0.50 & 616 & 6 \\
        $m_t$ \scriptsize{(\text{GeV})} & 168.26 & 0.75 & 168.27 & 0.018 & 168.28 & 0.028 & 168.27 & 0.01 \\
        $m_d$ \scriptsize{(\text{MeV})} & 2.67 & 0.19 & 2.58 & 0.48 & 2.59 & 0.44 & 2.58 & 0.03 \\
        $m_s$ \scriptsize{(\text{MeV})} & 53.16 & 4.61 & 54.33 & 0.25 & 54.81 & 0.36 & 54.29 & 0.53 \\
        $m_b$ \scriptsize{(\text{GeV})} & 2.839 & 0.026 & 2.841 & 0.075 & 2.848 & 0.350 & 2.839 & 0.015 \\
        $M_{D_1}$ \scriptsize{(\text{GeV})} & \multicolumn{2}{c}{---} & 8677 & - & 8790 & - & 8801 & 219 \\
        $M_{D_2}$ \scriptsize{(\text{GeV})} & \multicolumn{2}{c}{---} & 9724 & - & 9824 & - & 9820 & 189 \\
        $M_{D_3}$ \scriptsize{(\text{GeV})} & \multicolumn{2}{c}{---} & 96687 & - & 96710 & - & 96673 & 71 \\
        $\sin(\theta_{12})$ & 0.22650 & 0.000431 & 0.22653 & 0.071278 & 0.22671 & 0.48643 & 0.22649 & 0.000235 \\
        $\sin(\theta_{23})$ & 0.04053 & $^{+0.000821}_{-0.000601}$ & 0.04066 & 0.18581 & 0.04086 & 0.47030 & 0.04066 & 0.000292 \\
        $\sin(\theta_{13})$ & 0.00361 & $^{+0.000110}_{-0.000090}$ & 0.00359 & 0.15861 & 0.00364 & 0.30446 & 0.00359 & 0.000036 \\
        $\sin(\theta^{K^\pm}_{12})$ & \multicolumn{2}{c}{---} & 0.67511 & - & 0.67502 & - & 0.67497 & - \\
        $\sin(\theta^{K^\pm}_{23})$ & \multicolumn{2}{c}{---} & 0.06082 & - & 0.06112 & - & 0.04728 & - \\
        $\sin(\theta^{K^\pm}_{13})$ & \multicolumn{2}{c}{---} & 0.00721 & - & 0.00693 & - & 0.00152 & - \\
        $\sin(\theta^{K^{0}}_{12})$ & \multicolumn{2}{c}{---} & 0.82460 & - & 0.82462 & - & 0.82438 & - \\
        $\sin(\theta^{K^{0}}_{23})$ & \multicolumn{2}{c}{---} & 0.03327 & - & 0.03345 & - & 0.07482 & - \\
        $\sin(\theta^{K^{0}}_{13})$ & \multicolumn{2}{c}{---} & 0.02290 & - & 0.02335 & - & 0.02857 & - \\
        \hline
        $\chi^2$ & \multicolumn{2}{c}{---} & \multicolumn{2}{c}{$0.777$} & \multicolumn{2}{c}{$1.491$} & \multicolumn{2}{c}{$1.895$} \\
        \hline\hline
        \end{tabular}
        }
   \caption{Model various benchmark points with smallest $\chi^2$, smallest $\sigma_{\text{max}}$, and mean value for $\forall \sigma_{\text{max}}\leq 1$. Here $\sigma$ stands for standard deviation and has no units. Obtained values shown above were rounded to have the same significant figures as experiment results.}
    \label{tab:benchmark_obs}
\end{table}

The masses and mixing angles in Tab.~\ref{tab:benchmark_obs} were defined as eigenvalues of mass matrices in Eqs.~\eqref{eq:MassM_u},~\eqref{eq:MassM_dD}, and in Eq.~\eqref{eq:CKMangles} for $V^W_{CKM}$, $V^{K^{\pm}}$, $V^{K^0}$ mixing matrices, respectively.

Figure~\ref{fig:max_sigma_chi2} summarizes all the data points collected according to two criteria: horizontal axis corresponds to $\sigma_{\text{max}}$ which represents the maximum deviation of each point with respect to the experimental value, vertical axis shows the $\chi^2$ values for each point obtained. The plot is divided into three horizontal regions according to the value of $\sigma_{\text{max}}$: $0-1$, $1-2$, $2-3$; vertical region is separated into three categories as well, according to the values of $\left\langle\sigma\right\rangle / \sigma_{\text{max}}$: $0-0.4$, $0.4-0.6$, $0.6-1.0$. This last category represents the spread of all errors that contribute the total $\chi^2$. The solid curves on the plot stand for upper and lower theoretical limits for this plot given by $\chi^2=9\sigma_{\text{max}}^2$ and $\chi^2=\sigma_{\text{max}}^2$, respectively.
%
%
\section{Discussion}
\label{sec:discussion}
One can isolate and determine the causes of different levels of correlation between parameters and observable variables in the plots given in the earlier section, Figs.~\ref{fig:bg_observ_corr_A} and~\ref{fig:observ_corr_A}. Obviously, $\gamma$ and $\beta$ parameters affect mass values of quarks at up and down sector. Another factor in determining masses of the SM down sector quarks is the existence of BSM heavy isosinglet quarks, hereafter \emph{BSM effect}. For example, one would expect that $\gamma_u$ correlates with $m_u$ strongly, and $m_c$ and $m_t$ weakly. However, since $\beta_u$ is about 80 times greater than $\gamma_u$, strong correlation of $\gamma_u$-$m_u$ is smeared into medium level through interference of $\beta_u$. As expected,  $\beta_u$ correlates strongly with $m_c$ and $m_t$. Due to the relative size of $\beta_u$ with respect to $\gamma_u$, it affects  $m_u$ weakly.

However, the situation is different in down sector. Similar to the up sector, $\gamma_d$ correlates with  $m_d$ on the medium level. Correlation between $\gamma_d$ and other down quark mass eigenvalues ($m_s$ and $m_b$) disappears due to BSM effect. Correlation of $\beta_d$ with $m_d$, $m_s$, and $m_b$ is degraded proportional with the closeness to BSM quarks. Therefore, similarly to the up quark sector, weak correlation of $\beta_d$-$m_d$ have disappeared due to small BSM effect. Whereas, expected strong correlation between $\beta_d$ and $m_s$ weakens down to the medium level due to the existence of the same BSM effect. Finally, strong correlation of $\beta_d$-$m_b$ have disappeared due to very strong mixing with BSM quarks.

As mentioned earlier, CP violating phases are not considered in the present paper. Therefore, elements of mass matrices are chosen as real numbers. Consequently, some of the resulting eigenvalues of mass matrices and some elements of the CKM matrix are negative. It is possible to remove these negative signs and get correct CP violating phases by including phase multipliers to democratic mass matrix elements. These multipliers may even help to pinpoint $\chi^2$ and  $\sigma_{\text{max}}$. The affect of the phases on the quark masses and CKM mixing angles is left for the future works.
%
\section{Conclusion}
\label{sec:conclusion}
In the present paper, DMM approach is applied to the quark sector of the 331 model, which is inspired by $E_6$ symmetry. Model becomes prominent by being one of the simplest extensions of SM. Quark masses and mixing angles within one standard deviation of the experimental values with ten parameters are successfully derived. More specifically, each of the quark sectors(up, down, and isosinglet down) is controlled dominantly by set of three parameters ($a, \beta, \gamma$), additionally one parameter corresponds to the mixing between SM and isosinglet down type quarks. In return, all SM and isosinglet quark masses and mixing angles are predicted, total of eighteen observable variables nine out of which are SM variables. 

Detailed analysis is performed in order to find the best fit benchmark point. The best fit point obtained has a $\chi^2=0.777$ with the largest standard deviation from the experimental values of $0.586$ for $m_u$. Another important benchmark point is the point with the smallest achieved standard deviation error from the experimental data has $\chi^2=1.491$ and the largest deviation of $0.501$. Furthermore, a summary data plot of $\sigma_{\text{max}}$ vs $\chi^2$ is produced, as well as, the average point for all generated data with $\sigma_{\text{max}}\leq1$ condition.

The model at hand demonstrates that a democratic approach can successfully lead to the SM quark masses and hierarchy between them. Furthermore, CKM mixing angles are also obtained within corresponding experimental limits. This result motivates for further exploration on parameter schemes based on fundamental democratic pattern. UV models of flavour symmetry leading naturally to democratic based quark sector mass scheme should be studied in the future works. From all of the above, it is concluded that this can be a possible explanation of hierarchy problem.
%
%
\begin{acknowledgments}
RC was supported by Ege University Scientific Research Projects Coordination under Grant Number FGA-2021-22954. OP was supported by the Samsung Science and Technology Foundation under Grant No. SSTF-BA1602-04 and National Research Foundation of Korea under Grant Number 2018R1A2B6007000.
\end{acknowledgments}



\bibliography{references}
\end{document}